**Downburst Prediction Applications of GOES over the Western United States**

KENNETH L. PRYOR[1], STEVEN D. MILLER[2]

[1]NOAA/NESDIS/Center for Satellite Applications and Research, College Park, Maryland

[2]Cooperative Institute for Research in the Atmosphere, Fort Collins, Colorado

Corresponding author address:  Mr. Kenneth Pryor, Satellite Meteorology and Climatology Division, Operational Products Development Branch, NOAA/NESDIS/E/RA, NCWCP, Rm. 2833, 5830 University Research Court, College Park, MD 20740

E-mail: Ken.Pryor@noaa.gov

ABSTRACT

Over the western United States, the hazards posed to aviation operations by convective storm-generated downbursts have been extensively documented. Other significant hazards posed by convective downbursts over the intermountain western U.S. include the rapid intensification and propagation of wildfires and the sudden generation of visibility-reducing dust storms (haboobs). The existing suite of GOES downburst prediction algorithms employs the GOES sounder to calculate potential of occurrence based on conceptual models of favorable environmental thermodynamic profiles for downburst generation. Previous research has demonstrated the effectiveness of the Dry Microburst Index (DMI) as a prediction tool for convectively generated high winds. A more recently-developed diagnostic nowcasting product, the Microburst Windspeed Potential Index (MWPI) is designed to diagnose attributes of a favorable downburst environment: 1) the presence of convective available potential energy (CAPE), and 2) the presence of a deep surface-based or elevated mixed layer with a large temperature lapse rate. This paper presents an updated assessment of the MWPI algorithm, case studies demonstrating effective operational use of the MWPI product, and recent validation results. MWPI data were collected for downburst events that occurred during the 2014 convective season and were validated against surface observations of convective wind gusts as recorded by wind sensors in high-quality mesonetworks. Favorable validation results include a statistically significant correlation ($r > 0.6$) and low mean error ($< 1$ kt) between MWPI values and confirmed downburst wind speeds measured in situ.

## 1. Introduction

Over the intermountain western United States, and especially over the Great Basin region, convective-storm generated winds are the most prevalent of all severe convective weather types (i.e. hail, tornadoes, wind). Convective windstorms are often related to downbursts, defined in general as strong downdrafts that induce an outflow of potentially damaging winds on or near the ground (Fujita 1985, Wakimoto 1985). Over the western U.S., the hazards posed to aviation operations by convective storm-generated downbursts have been extensively documented. Since 1994, the National Transportation Safety Board (NTSB) has documented fifteen downburst-related aircraft accidents over the intermountain western U.S. alone, four of which involved fatalities. Other significant hazards posed by downbursts over the intermountain region include the rapid intensification and propagation of wildfires, and the generation of convective wind-driven dust storms, also referred to as haboobs. Miller et al. (2008) describes the process by which the previously identified environmental factors for downbursts often result in haboob generation, presenting results from an intensive field study over the Arabian Peninsula (UAE[2]) that was conducted during August and September 2004. Their results include a haboob modeling simulation, through which it was shown that seasonal haboob activity may account for a significant component of the regional transportable aerosol mass. The case studies presented in Miller et al. (2008) demonstrate the utility of data from remote sensing platforms such as satellite, lidar, and radar, and in situ radiometer and surface observations, in the detection and meteorological characterization of dust storms.

The Southwestern United States summer monsoon season [or North American monsoon (NA monsoon)] is particularly active with respect to convective storm activity. Typically, lower- and mid-tropospheric moisture, originating from the tropical Pacific Ocean, Gulf of California, and Gulf of Mexico, respectively, is transported northward and westward over the United States intermountain region where the afternoon boundary layer is deep and well-mixed with a temperature lapse rate approaching dry adiabatic (Adams and Comrie 1997). The migration and placement of a subtropical anticyclone, as well as the evolution of pre-existing convective storm activity over the Gulf of California region, influence the transport of both low and mid-level moisture into the southwestern United States.  Maddox et al. (1995) discussed three synoptic-scale patterns associated with severe thunderstorms over Arizona during the summer monsoon season: 1) the "Four Corners High" (Type I) resulting from a northward and westward shift in the subtropical ridge position to the Arizona/Utah border, 2) the "Great Basin High" pattern (Type II) characterized by a westward and northward shift of the subtropical ridge to western Utah and Nevada., and 3)  the "Trapping High" in which the subtropical ridge is depressed southward to the United States – Mexico border.

Another monsoonal phenomenon that influences convective storm development and incipient downburst generation is the "Gulf Surge" — the process by which lower tropospheric moisture is transported northwestward through the Gulf of California into the Sonoran Desert and Great Basin (Hales 1972).  Hales (1972, 1975) and Douglas and Li (1996) noted that the Gulf Surge is detectable in both radiosonde observation (RAOB) thermodynamic profiles and surface observation analyses over southern Arizona and southern California as a resultant increase in the southerly component of low-level winds and lower and middle tropospheric moisture, and a resultant decrease in surface temperatures.  Another result of the onset of the Gulf Surge is the preconditioning for convective instability and subsequent downburst generation that entails the establishment of a lower tropospheric moist air layer, originating from the GOC, underlying a middle-upper tropospheric dry air layer originating from the eastern North Pacific Ocean.  Intense solar heating over the Sonoran and Mojave Deserts and the Great Basin region, where the Bowen ratio (Lewis 1995) is typically very large, results in a vertical juxtaposition of the lower tropospheric moist layer and a well-mixed surface-based convective boundary layer in which relative humidity decreases with approach to the surface. This pattern fosters the presence of the inverted V thermodynamic profile that dominates the pre-convective environment featured in the case studies of this paper.

This study addresses recent modification, validation, and application of an algorithm that capitalizes on improvements to the Geostationary Operational Environmental Satellite (GOES) sounding process (Li et al. 2008) to extend the predictability of downburst-generated winds, especially to drier environments over the intermountain western United States. Schmit et al. (2008) compare the GOES-R Advanced Baseline Imager (ABI) sounding profile to previous versions available from GOES-13 through 15, stating that ABI spatial resolution and refresh rate will be twenty times greater with much finer spatial coverage. In addition, temperature profile accuracy will be slightly improved when compared to numerical model forecast soundings. Moisture profile accuracy will be significantly greater than numerical model accuracy, especially for the 700 to 850 mb layer where background error will be up to 30% lower for ABI soundings. Thus, the ABI moisture profile in the warm-season sub-cloud layer will be highly reliable. In summary, Schmit et al. (2008) stated that "both the ABI and the current sounder provide slightly improved temperature information to the forecast…the ABI error is similar to that of the current GOES sounder. Both have three broad water vapor spectral bands; both instruments improve the forecast information that shows a large water vapor background error" The ABI will also have comparable instrument noise values to the current sounder and will produce a CAPE measurement, crucial for determining the potential for deep convective storms and resulting downburst generation.

The case studies presented in this paper represent noteworthy downburst events that occurred on the periphery of the NA monsoon domain over the southwestern United States during the summer of 2014 and will be discussed in terms of prevailing monsoon synoptic patterns. In addition to thermodynamic and cloud microphysical factors that result in downburst generation, the role of topography in channeling and enhancing convective outflow is also considered. A combination of the Venturi Effect, as governed by Bernoulli's principle and the continuity equation (Halliday et al. 1993), and gravitational force acting on thunderstorm outflow likely result in higher measured wind gust speeds associated with downbursts that occur in mountainous regions.

The suite of GOES-derived downburst prediction algorithms employs data from the Sounder instrument to calculate risk based on conceptual models of favorable environmental thermodynamic profiles for downburst occurrence (Pryor 2015). Previous research efforts (Ellrod et al. 2000, Pryor and Ellrod 2004) have demonstrated the effectiveness of the Dry Microburst Index (DMI) as a prediction tool for the occurrence of thunderstorm-generated high winds:

$$DMI \equiv \Gamma + (T\text{-}T_d)_{700} - (T\text{-}T_d)_{500} \qquad (1)$$

where $\Gamma$ represents the absolute value of temperature lapse rate (negative vertical gradient in ambient temperature) between the 500 and 700 mb levels, and $(T\text{-}T_d)$ represents the dewpoint depression at two different levels of the atmosphere. As defined, DMI returns significant positive values for conditions of strong lapse rate (favoring dry adiabatic conditions conducive to evaporation) with additional weight given to moist-over-dry profiles (the so-called 'inverted V' thermodynamic profile). The authors note that the DMI should be $\geq 8$ for dry downbursts to occur, and that no correlation has been established between DMI values and maximum wind speeds associated with downbursts. A limitation of the DMI is the lack of a predictor that accounts for updraft intensity and associated precipitation loading. Pryor (2015) states the importance of the CAPE parameter for inferring updraft intensity and resulting precipitation type and concentration within a convective storm, which subsequently influences downdraft intensity. Previous GOES-based convective cloud studies (Lindsey et al. 2006), conducted in similar thermodynamic environments over the western United States, emphasize the importance of large CAPE in generating a volume of a large number of small precipitation particles. A high concentration of smaller mixed-phase precipitation particles within a convective storm will typically produce stronger downdrafts (Srivastava 1987).

In order to extend the utility of GOES sounder-derived data to multiple regions, Pryor (2014, 2015) outline the relevant physical processes and associated parameters deemed most important for downburst occurrence in intermediate thermodynamic environments over the central and eastern United States, and formulating the Microburst Windspeed Potential Index (MWPI):

$$\text{MWPI} \equiv \{(\text{CAPE}/1000 \text{ J kg}^{-1})\}+\{\Gamma/5 \text{ °C km}^{-1}+((\text{T-T}_d)_{850}-(\text{T-T}_d)_{670})/5 \text{ °C}\} \qquad (2)$$

Pryor (2014, 2015) show that the MWPI, an expansion of the DMI formula in which the calculation boundaries (670 and 850 mb levels) are located at a lower level in the troposphere, effectively identifies lower tropospheric thermodynamic structures favorable for downburst formation. According to Maddox et al. (1995), for the Type I pattern, "The afternoon boundary layer is nearly dry adiabatic to 670 mb, presenting a very favorable environment for downbursts and strong outflows." This statement underscores the importance of the selection of 670 mb as the upper boundary for the MWPI calculation, especially in summer monsoon convective environments. In addition, for the western U.S., functionality has been implemented in the downburst prediction algorithm that employs an alternate computation for temperature lapse rate and dewpoint depression difference for the $500 - 700$ mb layer over the higher elevations of the intermountain region:

$$\text{MWPI} \equiv \{(\text{CAPE}/1000 \text{ J kg}^{-1})\}+\{\Gamma/5 \text{ °C km}^{-1}+((\text{T-T}_d)_{700}\text{-(T-T}_d)_{500})/5 \text{ °C}\} \qquad (3)$$

MWPI improves upon the DMI product, providing a correlation of greater than 0.6 between index values and downburst-related wind gust speeds with a confidence level near 100%. A study by Caplan et al. (1990), in which afternoon sub-cloud temperature lapse rates are found to be correlated with downburst activity, validates the choice of the $500 - 700$ mb layer in the short-term prediction of downburst potential. Caplan et al. (1990) presented results derived from Joint Airport Weather Studies (JAWS) (McCarthy et al. 1982) and the Mesogamma 85 field experiment that indicated a functional relationship between $500 - 700$ mb temperature lapse rates and downburst occurrence, with a lapse rate of $5°\text{C km}^{-1}$ identified as the minimum threshold.

Responding to growing interest in satellite-based severe convective storm nowcasting in anticipation of GOES-R, this paper will present initial results of an experimental blending of the MWPI and the University of Alabama Huntsville (UAH) convective initiation (CI) algorithm, and explore the contribution of particular CI interest fields, such as the 6.5–10.7-μm brightness temperature difference (BTD). The GOES Satellite Convection Analysis and Tracking (SATCAST) system (Walker et al. 2012, Mecikalski et al. 2015) has recently demonstrated success in short-term forecasting of convective storm development. The most recent version of SATCAST uses nine satellite predictors and 16 numerical weather prediction (NWP) model predictors, and then applies a logistic regression (LR) modeling technique to produce a two-hour probabilistic forecast of deep moist convection. A favorable outcome of the implementation of the latest version of SATCAST is a reduction in false alarm rates and an improvement in skill. The benefits expected by implementing the MWPI algorithm in the SATCAST system is twofold: a reduction in MWPI false alarm rates, and extension of the SATCAST system to severe convective storm prediction, especially the short-term prediction of hazardous winds.

## 2. Validation Process and Results

The process for this research effort follows the methodology outlined in Pryor (2015). MWPI algorithm output data were collected for downburst events that occurred over the Great Basin region during the summers (June to September) of 2013 to 2015 and were validated against measured surface observations of convective wind gusts. Over the western U.S., the Great Basin was selected as the region of study for the following reasons: 1) topographic diversity that includes mountain ranges and valleys adjacent to larger expanses of arid lowlands, 2) summer convective storm activity related to the NA monsoon, and 3) the presence of wind sensors in high-quality mesonetworks over southern Idaho, western Utah, Nevada, and southern California. Mesonetworks sampled for this study and available via MesoWest (http://mesowest.utah.edu) include U.S. Bureau of Reclamation AgriMet, University of Utah MesoWest Mesonet (UUNET), U.S. Army Dugway Proving Ground (DPG), Desert Research Institute (DRI), Idaho Transportation Department (ITD), Nevada Department of Transportation (NVDOT), National Weather Service/Reno Weather Forecast Office (REVWFO), NOAA Air Resources Laboratory Field Research Division (ARL FRD) and Special Operations and Research Division (ARL SORD), and Remote Automatic Weather Stations (RAWS).

To intercompare the performance of the MWPI to legacy downburst prediction techniques, CI probability datasets were obtained from UAH and displayed over DMI, downdraft CAPE (DCAPE), and MWPI product images for case studies of two high-impact downburst events that occurred on 1 July and 18 August 2014. MWPI and DMI values were calculated from GOES-15 sounding retrieval datasets. Since DCAPE is not calculated as a GOES sounder derived product, this parameter was calculated from Rapid Refresh (RAP) model datasets using the formula (Eq. 1) as documented in Gilmore and Wicker (1998). For intercomparison purposes in each case study, the downdraft source height in the DCAPE calculation was set to equal the upper boundary of the MWPI calculation. The conversion of DCAPE values to wind gust potential applies the conservation of energy principle and work-energy theorem represented by the formula $W_{max} = (2\ DCAPE)^{1/2}$. As a supplement to the CI product, the brightness temperature difference (BTD) between the GOES-15 imager 6.5 μm and 11 μm channels (Schmetz et al. 1997) , was calculated and displayed with a color enhancement for each case study. Positive  BTD is associated with  a deep convective storm cloud top that has reached the height of the tropopause and lower stratosphere.  A threshold BTD of > −10°C indicates the presence of mature convective clouds, with tops approaching the upper troposphere that will soon likely generate precipitation (Mecikalski and Bedka 2006). For the 18 August 2014 case, in which the BTD signal for deep convection was weak, the BTD time difference ($\partial BTD / \partial t$) was calculated and displayed to show convective cloud growth rates that can be associated with convective storm initiation. The time difference calculation assumed slow translational motion during a short time period (< 15 min). GOES-15 imager datasets were obtained from the NOAA Comprehensive Large Array-data Stewardship System (CLASS; http://www.nsof.class.noaa.gov).

As previously discussed, the influence of the NA Monsoon is conducive to severe thunderstorm outbreaks that occur in the Great Basin region during the summer.  The NOAA/Storm Prediction Center (SPC) severe thunderstorm event database (available at http://www.spc.noaa.gov/wcm/#data) indicates that during the period of 1955 to 2014, wind reports of 26 m s$^{-1}$ (50 kt) or greater constituted the majority (>50 %) of all severe thunderstorm event reports over the intermountain western U.S., with an average of 144 severe wind events per year.  Among these events 80% occurred during the monsoon season, considered to be 1 June to 30 September for the purpose of this study. The SPC database also shows that the convective seasons of 2013 and 2014 were quite active compared to the long-term average, with the majority of the events recorded in 2013 (459 of 482) and in 2014 (294 of 338) occurring during the monsoon season. The large number of convective wind reports that occur during the monsoon season underscores the motivation for a downburst prediction algorithm validation study over the western U.S.

To conduct this 2013/2014 monsoon season validation, the prevailing weather pattern was identified by overlying 13-km resolution Rapid Refresh (RAP) model (available at http://nomads.ncdc.noaa.gov/data.php) analysis 500 mb geopotential height contours over GOES-15 water vapor imagery for each recorded downburst event. Table 1 summarizes the validation results for the Great Basin region. Favorable results include a positive correlation (r = 0.58), with a confidence level near 100%, between MWPI values and downburst wind gust magnitude. Accuracy was evaluated by calculating the mean forecast error (MFE) and mean absolute error (MAE) between MWPI-derived wind gust predictions and observed downburst-related wind gust speeds in accordance with the following equations:

$$\text{MFE} = \frac{\sum_{i=1}^{n} e_i}{n} \tag{4}$$

$$\text{MAE} = \frac{\sum_{i=1}^{n} |e_i|}{n} \tag{5}$$

Error (e) is defined as the difference between the observed wind gust speed and the predicted wind gust value based on the MWPI. MFE represents the bias of the MWPI wind gust prediction while MAE represents absolute error size, or accuracy. Especially encouraging is a very small MFE (or bias) of less than 0.001 kt that indicates the MWPI has no tendency to either underestimate or overestimate wind gust potential. An MAE of 5.5 kt calculated over the Great Basin region and an MAE of 4.6 kt over the western Great Basin region (California and Nevada) indicates, in general, that the MWPI is accurate to within five (5) kt of observed measured downburst wind gusts.  Isolating the western Great Basin region of Nevada and California in statistical computation also yielded more favorable results that include a slightly higher correlation of r = 0.61) for MWPI values calculated at GOES-15 sounder retrieval locations and a correlation of r = 0.64 for grid-interpolated MWPI values.  Pryor (2015) discusses sources of error for the MWPI that include precipitation phase and size distribution, storm translational motion, and the presence of a rear-inflow jet. Even more importantly, the larger error apparent in the scatterplot in Figure 1 likely results from terrain effects, as will be discussed in the July 2014 case study.

**3.  Case Studies**

   *a. 1 July 2014 Lake Tahoe-Reno Area Downbursts and Dust Storm*

   At the beginning of July 2014, the "Great Basin High" (Type II) pattern was well-established over the southwestern United States. The 2200 UTC 1 July 2014 GOES-15 water vapor image with overlying 500 mb geopotential height contours in Figure 2 shows the subtropical anticyclone positioned over the Nevada-Arizona border and a shortwave trough moving northeastward along the northwestern periphery of the anticyclone toward the Sierra Nevada range. During the early afternoon of 1 July, clusters of intense thunderstorms developed over western New Mexico and the Mogollon Rim of Arizona, which are typical convective initiation regions during the NA monsoon season. Mid-tropospheric southerly to southwesterly flow over the Sierra Nevada range resulted in lee-side  orographic initiation of deep convective storms, near the time of maximum surface heating, that tracked northward toward the Carson City and Reno, Nevada metropolitan areas. Over the western Great Basin region, the lower boundary layer was very dry with surface dew point temperatures below 0°C as shown in the RAOB sounding thermodynamic profile over Reno in Figure 3. The thermodynamic profile also shows a nearly 200 mb deep layer of mid-tropospheric moisture that likely originated from the subtropical eastern Pacific Ocean, the Gulf of Mexico, and from the redistribution of moisture by previously occurring deep convective storms within a shear zone over the lower Great Basin and Mogollon Rim regions.

As exemplified in Wakimoto (1985), intense surface heating through the daytime hours resulted in the development of a classic inverted V profile that supported the development of dry downbursts. Accordingly, the southwestern CONUS sector MWPI product image in Figure 2 shows index values of four or greater, indicating a wind gust potential greater than 23 m s$^{-1}$ (45 kt) over Sierra Nevada and most of the Great Basin region. Figure 4 shows the local-scale MWPI, DMI, and DCAPE product over the Reno-Lake Tahoe area at 2200 UTC. Over the immediate Reno-Lake Tahoe area, MWPI values were calculated for the 500 to 700 mb layer while farther north and east, over the lower elevations of the Great Basin, the MWPI was calculated for the 670 to 850 mb layer, similar to the calculation over central and eastern CONUS. In addition Figure 4, reveals slightly higher MWPI values (> 5) resulting from the 670–850 mb calculations. Higher 670–850 mb MWPI values may result from the presence of a superadiabatic surface layer caused by intense solar heating, which would increase the value of the temperature lapse rate parameter. Convective storm probability greater than 30 percent (light blue shading) was indicated near the California-Nevada border in a region with a secondary local maximum in MWPI values. MWPI values between 4.5 and 5, extending from Lake Tahoe northward along the California-Nevada border indicated wind gust potential of 26 to 27 m s$^{-1}$ (50 to 53 kt) where strong storms eventually developed and produced a downburst wind gust of 26 m s$^{-1}$ (51 kt) at Doyle, California ("DYL") near 0200 UTC. Overlying BTD > −10°C (red to magenta shading) signified the intensity and maturity of convective storm activity immediately downstream of the Sierra Nevada range that was capable of triggering new development in the Carson City (location "DU") and Reno ("RNO") areas. The DMI and DCAPE products indicated local maxima farther east over central Nevada where convective storm probability was near zero. Maximum predicted surface downdraft velocities ($W_{max}$) calculated from DCAPE were greater than 40 m s$^{-1}$ (80 kt) over the Reno-Lake Tahoe area and well over 46 m s$^{-1}$ (90 kt) in the local maximum region over central Nevada. WV-IR BTD product imagery displayed in Figure 5 indicate rapid intensification of convection over the Sierra Nevada and expansion of the area of convective storms over the western Great Basin region. Especially noteworthy is the rapid intensification of convective storm activity in the Carson City – Reno area between 2300 and 2330 UTC as evidenced by the increase in areal coverage of BTDs approaching zero (magenta shading) and corresponding increase in areal coverage of high reflectivity factor convective storms (> 40 dBZ) extending from the Reno area northwestward into northeastern California. The appearance and growth of convection in Figure 5 agrees well with the diagnosis of the CI product for the development of storms with reflectivity factors greater than 35 dBZ within the following one to two hour period. Reno  NEXRAD  reflectivity  factor  imagery shown in Figure 6 indicated that by 2330 UTC three distinguishable clusters of convection had developed over the Sierra Nevada range near the California-Nevada border. At this time, the northernmost storm cell was moving northward along the eastern slopes of the Carson Range between Carson City and Little Valley, and contained a high reflectivity factor (> 50 dBZ) with a protrusion echo pointing westward toward Little Valley. At 2334 UTC, the storm cell structure had evolved into a concave bow echo pattern with outflow directed toward Little Valley. Less than five minutes later, near 2338 UTC, Little Valley Remote Automatic Weather Station (RAWS) recorded a wind gust of 35 m s$^{-1}$ (68 kt) from a southeasterly direction, most likely originating from the strong storm cell located about 7 km southeast of the station.

Near this time, observers noted "blowing dust reducing visibility to 0.8 km at times" in the eastern Carson City area along United States (US) Highway 50. The 2330 UTC Reno NEXRAD reflectivity factor imagery indicated that this dust storm was co-located with a protrusion echo pointing eastward over Carson City. Vertical transect displays from the Reno NEXRAD also provided evidence of downburst occurrence in the Carson City area. Comparison of reflectivity imagery to differential reflectivity factor ($Z_{DR}$) imagery (Figure 7) show that at a range near 70 km from the NEXRAD site (location marked "DU"), $Z_{DR}$ values near 0 dB (indicative of hail) extending from the melting level downward toward the lower boundary of the radar beam , are co-located with high $Z_h$ (> 45 dBZ). Wakimoto and Bringi (1988) noted that this unique signature, called a "$Z_{DR}$-hole", is associated with a downburst-producing convective storm in which a descending core of melting hail is being forced downward by an intense downdraft. This signature appeared in NEXRAD imagery between 2330 and 2334 UTC, just prior to the observation of a 35 m s$^{-1}$ (68 kt) wind gust at Little Valley RAWS and the dust storm in Carson City.

In addition to thermodynamic and cloud microphysical factors that resulted in a downburst, we must also consider the role of topography in enhancing convective outflow. A combination of the Venturi Effect, as governed by Bernoulli's principle and the continuity equation, and gravitational force acting on the thunderstorm outflow likely resulted in a higher measured wind gust speed at Little Valley RAWS than the wind gust potential speed indicated by the closest MWPI value. NEXRAD imagery between 2330 and 2334 UTC indicated that downburst generation occurred over the eastern slopes of the Carson Range about 4 km east of Little Valley RAWS. The thunderstorm downdraft and resulting outflow, with a lower temperature and increased density due to melting of hail and evaporation of rain, was likely forced and channeled westward through narrow mountain passes and then downslope by gravity into Little Valley. The wind gust of 35 m s$^{-1}$ (68 kt) from an easterly direction and the direction of the point of closest approach by the parent storm supports the reasoning of the influence of topography on the magnitude of downburst-related winds observed at Little Valley. Within an hour, as noted in Table 2, a downburst-related wind gust of 28 m s$^{-1}$ (55 kt) was recorded at Slide Mountain mesonet (DRI) station, located on the summit at an elevation of 2.9 km and 7 km north of Little Valley RAWS. As the closest downwind observing station, this wind measurement was most suitable to compare to the observation at Little Valley RAWS to investigate the influence of topography. A convective storm, with similar reflectivity factor and structural characteristics (not shown) to the storm that affected Little Valley RAWS, and a member of the same complex of convective storms, passed over the Slide Mountain station between 0020 and 0040 UTC 2 July where terrain effects (i.e. friction, channeling) are assumed to be minimized. The observed wind gust of 28 m s$^{-1}$ (55 kt) at Slide Mountain was much closer to the WGP of 26 m s$^{-1}$ (51 kt) as documented in Table 2.

For this downburst event over the intermountain western U.S., MWPI values of 4.5 or greater were associated with measured wind gusts of 25 m s$^{-1}$ (50 kt) or greater . The CI probability algorithm, applied with the MWPI product, effectively indicated severe downburst potential over the greater Reno area and adjacent northeastern California for a one to four hour period following the 2200 UTC product generation time. This demonstrates the operational utility of a merged CI-MWPI product in which the highest storm probability was nearly co-located with a local maximum in wind gust potential.

*b. 18 August 2014 Southern California Downbursts*

During the afternoon of 18 August 2014, clusters of convective storms developed over the Mogollon Rim of Arizona, and over the San Bernardino and San Jacinto Mountains of southern California, and remained nearly stationary. Analysis of satellite imagery and Rapid Refresh (RAP) model sounding thermodynamic profiles also revealed that a Gulf Surge was in progress over southwestern Arizona and southeastern California. The only severe wind event west of the Rocky Mountains recorded by the National Weather Service/Storm Prediction Center on 18 August occurred near Daggett, California at Barstow-Daggett Airport. Figure 8 shows a synoptic overview of developing convective storm activity over the western U.S. during the afternoon of 18 August 2014. GOES-15 6.7 μm water vapor (WV) imagery, with overlying 500 mb geopotenial height contours, prominently displays the presence of a "Four Corners High" flanked by a cut-off low centered over the northern California Pacific Coast and a northwest-southeast oriented trough over southern California and the Arizona-Mexico border. Comparison of the RAP model sounding thermodynamic profiles between Daggett and Twentynine Palms in Figure 8 reveals that the 500 mb trough centered over the U.S.-Mexico border is resulting from lower tropospheric cooling and the depression of 1000–500 mb thickness values due to the Gulf Surge. Also apparent in the regional MWPI product image in Figure 8 is a line of isolated convective storms developing along the western boundary of the Gulf Surge, where MWPI values increase along and west of the boundary. The western boundary of the Gulf Surge, roughly co-located with the axis of the San Bernardino and San Jacinto Mountains, has characteristics similar to a dry line, for example: 1) surface convergence with a wind shift from a southeasterly to a westerly direction, 2) a marked temperature and dew point temperature gradient with a 10°C decrease in dew point from east to west across the boundary, and 3) a deeper and drier convective boundary layer along and west of the Gulf Surge boundary. These factors, more typical of the dry line observed over the Great Plains, promoted convective storm initiation and subsequent strong downdraft generation.

Between 2000 and 2100 UTC, convective storms developed and intensified along the moisture discontinuity over the San Bernardino and San Jacinto Mountains. A comparison of local scale MWPI, DMI, and DCAPE product imagery in Figure 10 shows an increasing MWPI gradient between the California-Mexico border and the Mojave Desert, with the northernmost cluster of convective storms developing within this gradient between 2000 and 2100 UTC. The narrow zone of convective initiation with convection probability > 30%, extending from northern Baja California to the Nevada-California border, was characterized by increasing lower tropospheric instability from south to north as the boundary layer became deeper with a larger temperature lapse rate. Accordingly, at 2000 UTC, the closest representative MWPI value to Daggett of 5.5 corresponded to wind gust potential of 28 m s−1 (55 kt), while farther southeast at Twentynine Palms, in the moister air mass, the closest representative MWPI of 3.5 corresponded to wind gust potential near 23 m s−1 (45 kt). The DMI and DCAPE $W_{max}$ products indicated local maxima farther north and west over central California where convective storm probability was near zero. In a similar manner to the previous case, $W_{max}$ values were greater than 40 m s$^{-1}$ (80 kt) in the CI zone along the western boundary of the Gulf Surge and over 46 m s$^{-1}$ (90 kt) in the local maximum region over central California. The RAP model sounding profiles in Figure 9 illustrate the difference in wind gust potential over southern California. Although CAPE was higher over Twentynine Palms, which would favor thunderstorms with more intense precipitation (likely hail), the deeper dry adiabatic layer over Daggett supported greater downdraft energy in the sub cloud layer.

By 2200 UTC, a line of isolated innocuous-appearing convective storms was developing in the moisture discontinuity zone over southern California. Figure 11, the 2211 UTC WV-IR BTD product, displays values < -10°K that indicated early mature convective clouds that have not yet reached the tropopause. However, the 11-minute (2200 – 2211 UTC) WV-IR channel BTD time trend ($\Delta$BTD) indicated relatively large values (>20°K) associated with the storms developing near Daggett ("ĐAG") and Fort Irwin ("BYS"), signifying both movement and cloud top growth within the 15-minute period preceding the Daggett downburst. The large $\Delta$BTD values were a stronger signal for intensifying convective storms with the potential to generate heavy rainfall and severe downburst winds. Interestingly, the largest $\Delta$BTD values were found near the axis of maximum MWPI values extending from the Daggett area to Fort Irwin. The trend of storm intensification was still apparent at 2230 UTC, as illustrated in Figure 12. During the period from 2211 to 2230 UTC, significant horizontal and vertical growth of storms near Daggett and Twentynine Palms ("NXP") were apparent as expanding areas of upper-tropospheric cold cloud tops (BTD > -10°K) and large $\Delta$BTD values (>20°K).

Edwards Air Force Base (AFB) NEXRAD reflectivity factor Plan Position Indicator (PPI) imagery in Figure 13 displayed two prominent high reflectivity factor multicell storms over San Bernardino County: 1) near Daggett (DAG) that developed along the western boundary of the Gulf Surge, and 2) near Twentynine Palms (NXP) that developed along a northwest-southeast oriented convergence boundary north of the San Bernardino Mountains. The first downburst-related wind gust speed of 27 m s$^{-1}$ (52 kt) was recorded at Daggett Airport (DAG) at 2215 UTC. Figure 13 shows that at 2212 UTC the multicell storm near Daggett had elongated with a southwest-northeast orientation. Meanwhile, the southernmost cell, closest to the moisture boundary, appeared to be most intense with a maximum reflectivity factor greater than 55 dBZ. Over the next 20-minute period, a new cell developed and intensified on the southern end of the storm, and produced a second, weaker downburst at Daggett Airport, with a wind gust speed of 22 m s$^{-1}$ (42 kt) measured at 2238 UTC.

Coincident with the evolution of the storm over Daggett, the multicell storm in the Twentynine Palms area exhibited a similar evolution, with a northwest-southeast orientation and new cell development on the southeastern (upwind) flank of the storm. A downburst-related wind gust speed of 20 m s$^{-1}$ (39 kt) was recorded at Twentynine Palms Marine Corps Air-Ground Combat Center (NXP) at 2234 UTC. Similar to the downburst at Daggett, a reflectivity factor greater than 55 dBZ was observed with the southernmost cell within the Twentynine Palms storm. However, wind gust magnitude was somewhat lower at Twentynine Palms than at Daggett, despite the similar appearance between the two multicell storms.

Inspection of vertical transect imagery near the time of downburst occurrence at Daggett reveals an alternative signature to that discussed in the 1 July 2014 Carson City, Nevada case. Although reflectivity factor imagery indicated in both cases the presence of tall, intense cells with high values of reflectivity factor (> 40 dBZ; see Figs. 7 and 14) extending well above the freezing level, there was a marked difference in the appearance of vertical transect $Z_{DR}$ imagery over Daggett around the time of downburst, as shown in Figure 14. Namely, the 2212 UTC transect imagery displayed $Z_{DR}$ between 0 and 1 dB near the melting level, overlying higher $Z_{DR}$ between 3 and 5 dB from an altitude near 4 km MSL to the surface. This sharp vertical gradient in $Z_{DR}$ between the melting level and the surface was noted in Pryor (2015) and associated with the rapid melting of hail with the descending precipitation core in a high reflectivity factor convective storm. The combination of the rapid melting of hail and subsequent evaporation of rain in the deep, well mixed sub cloud layer likely enhanced downdraft intensity over Daggett and resulted in the observance of wind gust speeds greater than 26 m s$^{-1}$ (50 kt) at Daggett Airport at 2215 UTC. The second, weaker downburst at Daggett was associated with a generally lower reflectivity factor and a weaker vertical $Z_{DR}$ gradient between the melting level and the surface as shown in transect imagery at 2236 UTC in Figure 14. With this downburst event over southern California, a departure of the thermodynamic profile from the classic inverted V was seen to correlate with a decrease in MWPI values and weaker measured downburst-related winds at the surface. In addition, storm microphysics, specifically precipitation content and type (i.e. hail vs. rain) influenced downdraft intensity with individual cells that tracked over Daggett Airport between 2200 and 2240 UTC.

## 5. Discussion and Conclusion

Pryor (2015) presented case studies of wet-type downburst events occurring over the eastern United States. The case studies presented in this paper highlight the adaptability of the MWPI algorithm to varying climatic and geographic regions in the intermountain western United States.  In particular, they highlight the flexibility of the MWPI prediction algorithm, to the extent that it can be refined or "tuned" based on the regional climatology.

As shown in Figure 1, MWPI values of 4.5 or greater over the western U.S. correspond to convective wind gust potential of 26 m s$^{-1}$ (50 kt) or greater. The relationship between MWPI values and measured convective wind gusts that differs between the eastern and western U.S. likely results from i) the thermodynamic environment, ii) physical characteristics of convective storms, and iii) the complex terrain over the Great Basin region.  Srivastava (1987) found that a particle size distribution consisting of a high concentration of smaller particles, both ice and liquid, further enhances downdraft intensity.  The July 2014 Carson City downburst event exemplified convective winds that were generated in a very dry boundary layer with surface dew point temperature below 0°C.  Similar to conditions observed over the western High Plains, this very dry boundary layer promoted the development of convective storms composed of a high concentration of small ice-phase precipitation (i.e. graupel, small hail). In addition, terrain effects on thunderstorm outflow was also apparent as evidenced by one of the highest downburst-related wind gusts of the 2014 monsoon season recorded at Little Valley RAWS.

Comparison of the MWPI product to legacy downburst prediction techniques highlighted the importance of inclusion of a predictor for convective updraft strength and resultant precipitation content. As expected in drier convective environments, the DCAPE parameter greatly overestimated wind gust potential by up to a factor of two for both cases. In addition, local maxima in DCAPE, and derived wind gust potential, were displaced from regions with observed downburst-generated wind gusts. A similar tendency was also noted for the DMI product, where local maxima were biased toward clear-sky regions with very dry boundary layers where deep convective storms were unlikely as indicated by the CI probability product. Gilmore and Wicker (1998) noted a number of deficiencies in using DCAPE for convective wind gust prediction: 1) parcel theory assumptions used to compute DCAPE are significantly violated, 2) entrainment of environmental air dilutes thunderstorm downdrafts and significantly changes the $\theta_e/\theta_w$ of the parcels, 3) evaporative cooling rates required to maintain saturation within the strongest downdrafts did not occur in any of the authors' simulations, and 4) actual increases in kinetic energy due to evaporative cooling within the downdraft are much less than predicted by DCAPE. Observed wind measurements in the case studies of this paper echo the findings of Gilmore and Wicker as indicated by the overestimation of DCAPE-derived convective wind gust potential. These results underscore the importance of the CAPE parameter in defining convective precipitation potential and diagnosing the storm intensity necessary to promote strong downdrafts and subsequent outflow winds.

Results of the blending of the CI probability product with the MWPI were encouraging. In both case studies, the CI algorithm, applied with the MWPI product, effectively indicated severe downburst potential for a one to four hour period following the product generation time. In addition, two CI algorithm interest fields were considered for evaluation: the GOES-15 6.5 μm − 11 μm channel BTD and BTD time difference ($\partial BTD/\partial t$). Combined with the MWPI product, the BTD provided a stronger signal for convective storm development and intensification for the 1 July 2014 case, while the $\partial BTD/\partial t$ indicated a stronger signal for storm development for the 18 August 2014 case. This demonstrates the operational utility of a merged CI-MWPI product in which the highest storm probability was nearly co-located with a local maximum in wind gust potential.

A downburst prediction study over India currently in progress will highlight the international adaptability of the MWPI algorithm and explore the future use of Indian geostationary satellite data in the generation of downburst prediction products, especially in regions where the warm season is dominated by monsoon patterns. Preliminary results of the application of the MWPI product during summer 2015, derived from numerical weather prediction model data, shows promise over the Indian Peninsula region. Similar to the western United States, the MWPI and associated downburst intensity over India during the pre-monsoon and summer monsoon seasons are responsive to varying degrees of the inverted V thermodynamic profile, lower to middle tropospheric temperature lapse rates, and convective storm precipitation content.

*Acknowledgements*. The author thanks Dr. Laurie Rokke (Chief, Operational Products Development Branch, NOAA/NESDIS) and Dr. Xin-Zhong Liang (Graduate Advisor, Department of Atmospheric and Oceanic Science, University of Maryland, College Park), for their constructive comments and suggestions for this manuscript. The author also thanks Americo Allegrino and Jaime Daniels (NESDIS) for providing GOES sounding retrieval datasets

REFERENCES

Adams, D. K. and A. C. Comrie, 1997. The North American Monsoon. Bull. Amer. Meteor. Soc., 78, 2197–2213.

Caplan, S.J., A.J. Bedard, and M.T. Decker, 1990: The 700–500 mb Lapse Rate as an Index of Microburst Probability: An Application for Thermodynamic Profilers. *Journal of Applied Meteorology*, **29**, 680–687.

Douglas, M. W., and S. Li, 1996: Diurnal variation of the lower-tropospheric flow over the Arizona low desert from SWAMP—1993 observations. *Mon. Wea. Rev.,* **124,** 1211–1224.

Ellrod, G.P., L. Bottos, J.P. Nelson, W.P. Roeder, M.R. Witiw, 2000: Experimental GOES Sounder Products for the Assessment of Downburst Potential. *Wea. Forecasting*, **15**, 527-542.

Fujita, T.T., 1985: The downburst, microburst and macroburst. SMRP Research Paper 210, University of Chicago, 122 pp.

Gilmore, M. S., and L. J. Wicker, 1998: The influence of midtropospheric dryness on supercell morphology and evolution. *Mon. Wea. Rev.,***126**, 943-958.

Hales, J. E., 1972: Surges of maritime tropical air northward over the Gulf of California. *Mon. Wea. Rev.,* **100,** 298–306.

Hales, J. E., 1975: A severe southwest desert thunderstorm: 19 August 1973. *Mon. Wea. Rev.,* **103,** 344–351.

Halliday, D., R. Resnick, and J. Walker, 1993: *Fundamentals of Physics.* John Wiley and Sons, 1130 pp.

Lewis, J. M., 1995: The story behind the Bowen Ratio. *Bull. Amer. Meteor. Soc.*, **76**, 2433–2443.

Li, Z., J. Li, W. P. Menzel, T. J.Schmit, J. P. Nelson III, J. Daniels, and S. A. Ackerman, (2008): GOES sounding improvement and applications to severe storm nowcasting. *Geophys. Res. Lett.*, **35**, L03806, doi:10.1029/2007GL032797.

Lindsey, D. T., D. W. Hillger, L. Grasso, J. A. Knaff, and J. F. Dostalek, (2006): GOES climatology and analysis of thunderstorms with enhanced 3.9-μm reflectivity. *Mon. Wea. Rev.*, **134**, 2342–2353.

Maddox, R., D. McCollum, and K. Howard, 1995: Large-scale patterns associated with severe summertime thunderstorms over central Arizona. *Wea. Forecasting*, 10, 763–778.

McCarthy, J., J. W. Wilson, and T. T. Fujita, 1982: The Joint Airport Weather Studies Project. *Bull. Amer. Meteor. Soc.*, **63**, 15–22.

Mecikalski, J. R., and K. M. Bedka, 2006: Forecasting convective initiation by monitoring the evolution of moving cumulus in daytime GOES imagery. *Mon. Wea. Rev.*, **134**, 49–78.

Mecikalski, J. R., J. K. Williams, C. P. Jewett, D. Ahijevych, A. LeRoy, and J. R. Walker, 2015: Probabilistic 0–1-h convective initiation nowcasts that combine geostationary satellite observations and numerical weather prediction model data. *J. Appl. Meteor. Climatol.*, **54**, 1039-1059.

Miller, S. D., A. P. Kuciauskas, M. Liu, Q. Ji, J. S. Reid, D. W. Breed, A. L. Walker, and A. A. Mandoos, 2008: Haboob dust storms of the southern Arabian Peninsula, J. Geophys. Res., **113**, D01202, doi:10.1029/2007JD008550.

Pryor, K.L., and G.P. Ellrod, 2004: Recent improvements to the GOES microburst products. *Wea. Forecasting*, **19**, 582-594.

Pryor, K. L., 2014: Downburst prediction applications of meteorological geostationary satellites. *Proc. SPIE Conf. on Remote Sensing of the Atmosphere, Clouds, and Precipitation V*, Beijing, China, doi:10.1117/12.2069283.

Pryor, K. L., 2015: Progress and Developments of Downburst Prediction Applications of GOES. *Wea. Forecasting*, **30**, 1182–1200.

Schmetz, J., S. A. Tjemkes, M. Gube, and L. van de Berg, 1997: Monitoring deep convection and convective overshooting with METEOSAT. *Adv. Space Res.*, **19**, 433–441.

Schmit, T.J., J. Li, J. J. Gurka, M. D. Goldberg, K. J. Schrab, J. Li, and W. F. Feltz, 2008: The GOES-R advanced baseline imager and the continuation of current sounder products. *Journal of Applied Meteorology and Climatology*, **47**, 2696-2711.

Srivastava, R.C., 1987: A model of intense downdrafts driven by the melting and evaporation of precipitation. *J. Atmos. Sci.*, **44**, 1752-1773.

Wakimoto, R.M., 1985: Forecasting dry microburst activity over the high plains. *Mon. Wea. Rev.*, **113**, 1131-1143.

Wakimoto, R. M., and V. N. Bringi, 1988: Dual-polarization observations of microbursts associated with intense convection: The 20 July storm during the MIST project. *Mon. Wea. Rev.*, **116**, 1521–1539.

Walker, J. R., W. M. MacKenzie, J. R. Mecikalski, and C. P. Jewett, 2012: An enhanced geostationary satellite–based convective initiation algorithm for 0–2-h nowcasting with object tracking. *J. Appl. Meteor. Climatol.*, **51**, 1931–1949.

LIST OF FIGURE CAPTIONS

Figure 1. (Top) Geographic regions of interest within the continental United States (CONUS) for the validation of the MWPI algorithm. Locations of all severe wind reports from the NOAA/Storm Prediction Center database between 1955 and 2013 are plotted over the image (Courtesy of the NOAA/Storm Prediction Center. Available online at http://www.spc.noaa.gov/gis/svrgis/), and (bottom) scatterplot of MWPI values vs. measured downburst-related wind gust speeds (in kt) for 60 events over the Intermountain western U.S. between June 2013 and July 2015.

Figure 2. GOES-15 water vapor channel image with overlying RAP model analysis 500 mb geopotential height contours (in meters, top) and MWPI product image (bottom) at 2200 UTC 1 July 2014. MWPI contour interval is one (1) with contours for MWPI values of one (1), three (3), and five (5) labeled. White rectangular area shows the domain of the images in Figures 4 and 5.

Figure 3. RAOB sounding thermodynamic profile over Reno, Nevada at 0000 UTC 2 July 2014. The dashed curve in thermodynamic profile represents virtual temperature. "ML" marks the height of the melting level (0°C isotherm). Mean layer CAPE, represented by the shaded area, is in units of J $\text{kg}^{-1}$. Vertical axis is labeled in units of pressure (millibars) and horizontal axis is labeled in units of temperature (°C).

Figure 4. Comparison of GOES MWPI product (top), DMI product (middle), and RAP model DCAPE product (bottom) at 2200 UTC 1 July 2014. Contours of grid-interpolated MWPI and DMI values are overlying index values plotted at sounding retrieval locations and GOES-15 visible imagery (top, middle). Contours of grid-interpolated DCAPE values are overlying $W_{max}$ values (kt) plotted at RAP model grid points and GOES-15 visible imagery (bottom). The DMI contour interval is four (4) and the DCAPE contour interval is 100. Location identifiers of downburst occurrence are listed in Table 2.

Figure 5. Blended GOES-15 imager WV-IR BTD and MWPI (color contours) product, with overlying 0.5°elevation radar reflectivity factor (dBZ) from Reno, Nevada NEXRAD at at 2300 UTC (top) and 2330 UTC (bottom) 1 July 2014. Color bars for BTD are located in the upper-right corner of the top product image; radar reflectivity factor color bar is located at the bottom of the figure.

Figure 6.  Reno, Nevada 0.5° elevation NEXRAD reflectivity factor (dBZ) at 2330 UTC (top) and 2334 UTC (bottom) 1 July 2014 with 2200 UTC GOES-15 MWPI values plotted over the images.  Location identifiers of downburst occurrence are listed in Table 2. "DU" marks the location of the report of a dust storm near Carson City at 2335 UTC. MWPI color bar is shown in Figure 4.

Figure 7. Reno, Nevada NEXRAD base reflectivity factor  vertical transect images at 2330 UTC (top left) and 2334 UTC (top right) 1 July 2014; and differential reflectivity factor  vertical transect images at 2330 UTC (bottom left) and 2334 UTC (bottom right) 1 July 2014. "DU" marks the location of the report of a dust storm near Carson City at 2335 UTC. "ML" represents the height of the melting level indicated in the thermodynamic profile shown in Figure 3.  Black-circled region indicates the presence of a hail core near the surface associated with a "$Z_{DR}$–hole" signature.

Figure 8. GOES-15 water vapor channel image with overlying 500 mb geopotential height contours (in meters, top) and GOES-15 MWPI product image (bottom) at 2000 UTC 18 August 2014.  Contour interval is one (1) with contours for MWPI values of zero (0), two(2), and four (4) labeled.  White rectangular area shows the domain of the images in Figure 10.

Figure 9.  RAP model-derived sounding thermodynamic profile over Daggett at 2000 UTC (top) and over Twenty-Nine Palms at 2100 UTC (bottom) 18 August 2014. The dashed curve in thermodynamic profile represents virtual temperature. "ML" marks the height of the melting level (0°C isotherm). Mean layer CAPE, represented by the shaded area, is in units of J kg$^{-1}$. Vertical axis is labeled in units of pressure (millibars) and horizontal axis is labeled in units of temperature (°C).

Figure 10. Comparison of GOES MWPI product (top), DMI product (middle), and RAP model DCAPE product (bottom) at 2000 UTC 18 August 2014. Contours of grid-interpolated MWPI and DMI values are overlying index values plotted at sounding retrieval locations and GOES-15 visible imagery (top, middle). Contours of grid-interpolated DCAPE values are overlying $W_{max}$ values (kt) plotted at RAP model grid points and GOES-15 visible imagery (bottom). The DMI contour interval is four (4) and the DCAPE contour interval is 100. "BYS", "DAG", "LAX, and "NXP" represent the locations of Fort Irwin, Daggett, Twenty-Nine Palms, and Los Angeles, California, respectively.

Figure 11. Blended GOES-15 imager WV-IR BTD and MWPI (color contours) product (top) and blended 11-minute ΔBTD and MWPI product (bottom) over southern California at 2211 UTC 18 August 2014, with overlying 0.5°elevation radar reflectivity factor (dBZ) from Edwards Air Force Base NEXRAD at 2212 UTC. Color bars for BTD and ΔBTD are located on the left hand side of the product image, radar reflectivity factor color bar is located at the bottom of the figure. "BYS", "DAG", "LAX, and "NXP" represent the locations of Fort Irwin, Daggett, Twenty-Nine Palms, and Los Angeles, California, respectively.

Figure 12. Blended GOES-15 imager WV-IR BTD and MWPI (color contours) product at 2241 UTC (top) and blended 19-minute ΔBTD and MWPI product (bottom) over southern California at 2230 UTC 18 August 2014, with overlying 0.5°elevation radar reflectivity factor (dBZ) from Edwards Air Force Base NEXRAD at 2236 UTC. Color bars for BTD and ΔBTD are located on the left hand side of the product image, radar reflectivity factor color bar is located at the bottom of the figure. "BYS", "DAG", "LAX, and "NXP" represent the locations of Fort Irwin, Daggett, Twenty-Nine Palms, and Los Angeles, California, respectively.

Figure 13. Edwards AFB, California 0.5° elevation NEXRAD reflectivity factor (dBZ) at 2212 UTC (top) and 2236 UTC (bottom) 18 August 2014 with 2000 UTC GOES-15 MWPI values plotted over the images. MWPI color scale is shown in Figure 10. "BYS", "DAG", "LAX, and "NXP" represent the locations of Fort Irwin, Daggett, Twenty-Nine Palms, and Los Angeles, California, respectively.

Figure 14. Edwards AFB, California NEXRAD base reflectivity factor vertical transect images at 2212 UTC (top left) and 2236 UTC (top right) 18 August 2014; and NEXRAD differential reflectivity factor vertical transect images at 2212 UTC (bottom left) and 2236 UTC (bottom right) 18 August 2014. Black tick marks the location of Barstow-Daggett Airport. "ML" represents the height of the melting level indicated in the thermodynamic profile shown in Figure 9. Black-circled region indicates the presence of a hail core near the melting level.

Table 1. MWPI validation statistics based on direct comparison between index values and measured downburst wind gusts. In the regression line equations, "x" represents the MWPI value while "y" represents predicted wind gust speed.

|  | Western U.S. (N=60) | NV – CA (N=24) |
|---|---|---|
| MFE (kt) | 0.0003 | -0.0004 |
| MAE (kt) | 5.5 | 4.6 |
| Correlation (r) | 0.58 | 0.61 |
| *t* value | 43.71 | 30.67 |
| Critical Value (P< 0.0005) | 3.46 | 3.77 |
| Confidence Level | >99.95% | >99.95% |
| Regression line equation | y=5.1532x+27.158 | y=4.6322x+30.127 |

Table 2. Measured wind gusts and associated downburst risk values for the 1-2 July 2014 Lake Tahoe-Reno downburst outbreak.

| Station | Time (UTC) | Wind Gust Speed (kt) | MWPI (500–700) | MWPI WGP (kt) |
|---|---|---|---|---|
| Little Valley (LVY) | 2338 | 68 | 4.6 | 51 |
| Mason Valley (RES1) | 2350 | 43 | 4 | 48 |
| Slide Mountain (SLI) | 0030 | 55 | 4.6 | 51 |
| Reno Airport (RNO) | 0055 | 59 | 4.6 | 51 |
| Reno NWSO (REV) | 0111 | 54 | 4.6 | 51 |
| Doyle (DYL) | 0211 | 51 | 4.6 | 51 |

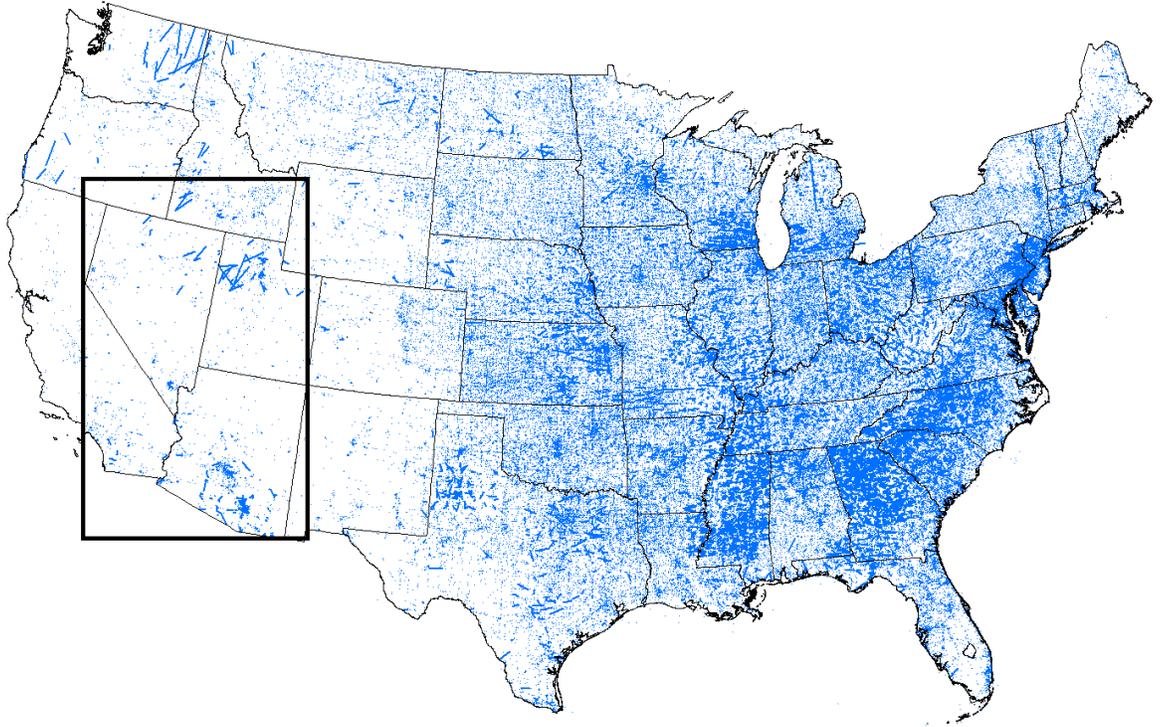

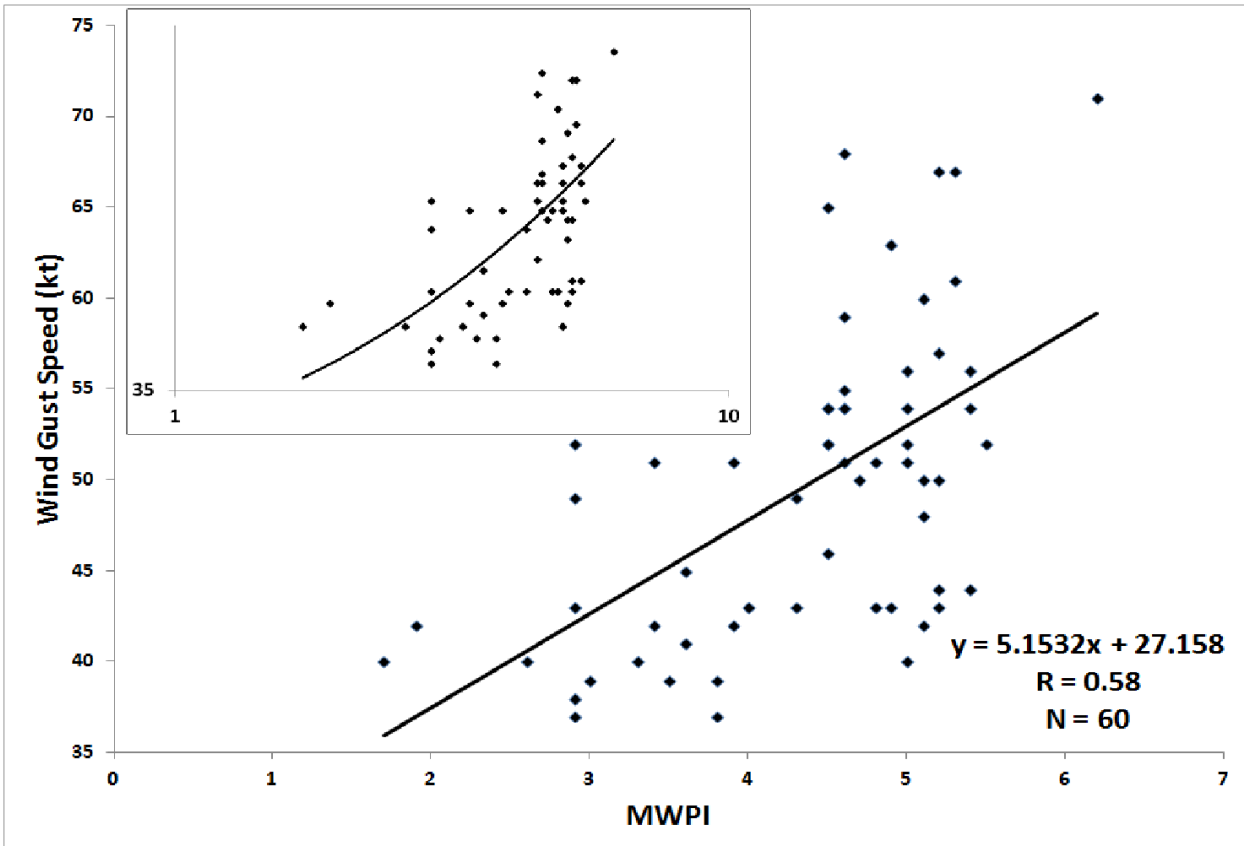

y = 5.1532x + 27.158
R = 0.58
N = 60

Figure 1. (Top) Geographic regions of interest within the continental United States (CONUS) for the validation of the MWPI algorithm. Locations of all severe wind reports from the NOAA/Storm Prediction Center database between 1955 and 2013 are plotted over the image (Courtesy of the NOAA/Storm Prediction Center. Available online at http://www.spc.noaa.gov/gis/svrgis/), and (bottom) scatterplot of MWPI values vs. measured downburst-related wind gust speeds (in kt) for 60 events over the Intermountain western U.S. between June 2013 and July 2015.

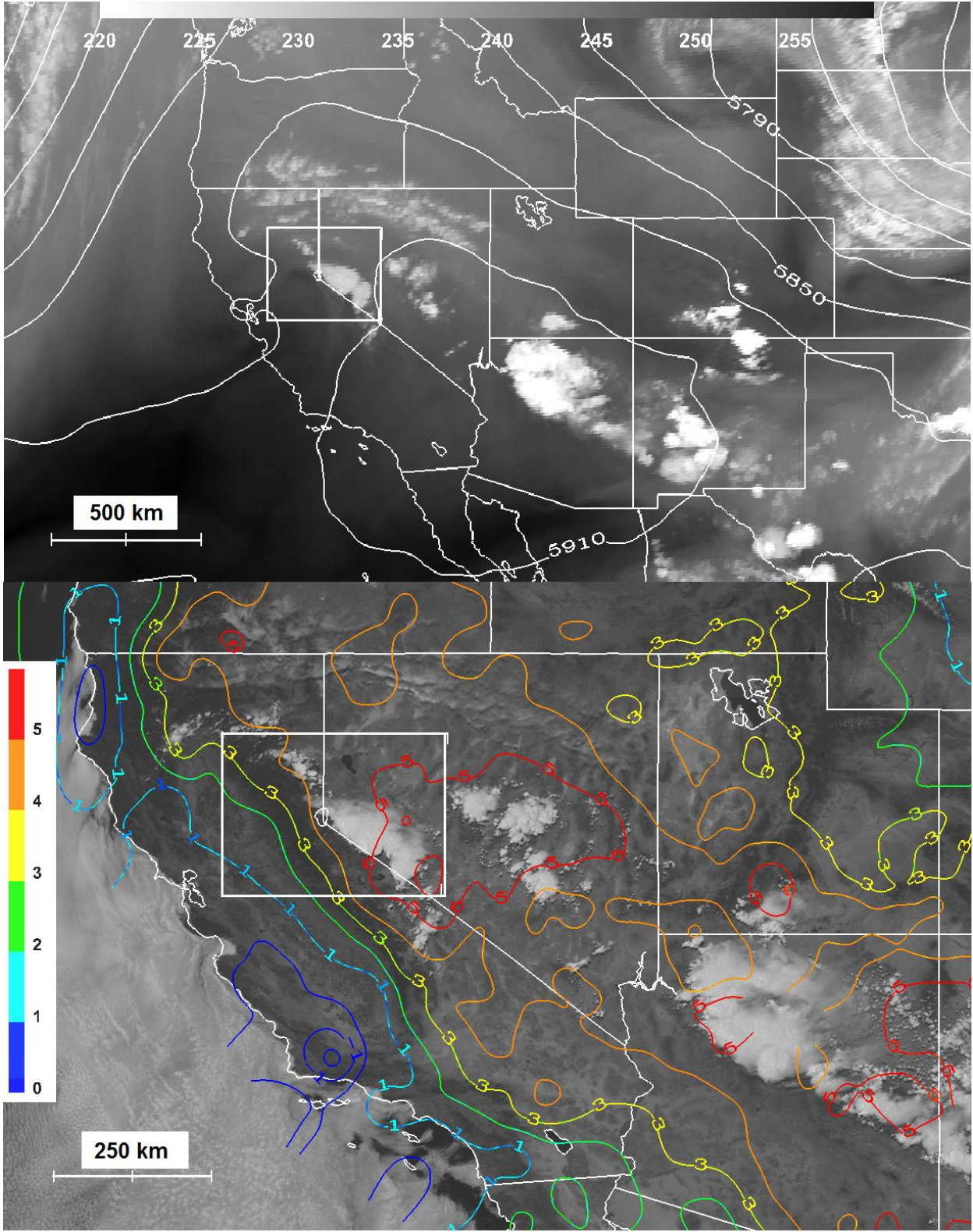

Figure 2. GOES-15 water vapor channel image with overlying RAP model analysis 500 mb geopotential height contours (in meters, top) and MWPI product image (bottom) at 2200 UTC 1 July 2014. MWPI contour interval is one (1) with contours for MWPI values of one (1), three (3), and five (5) labeled.  White rectangular area shows the domain of the images in Figures 4 and 5.

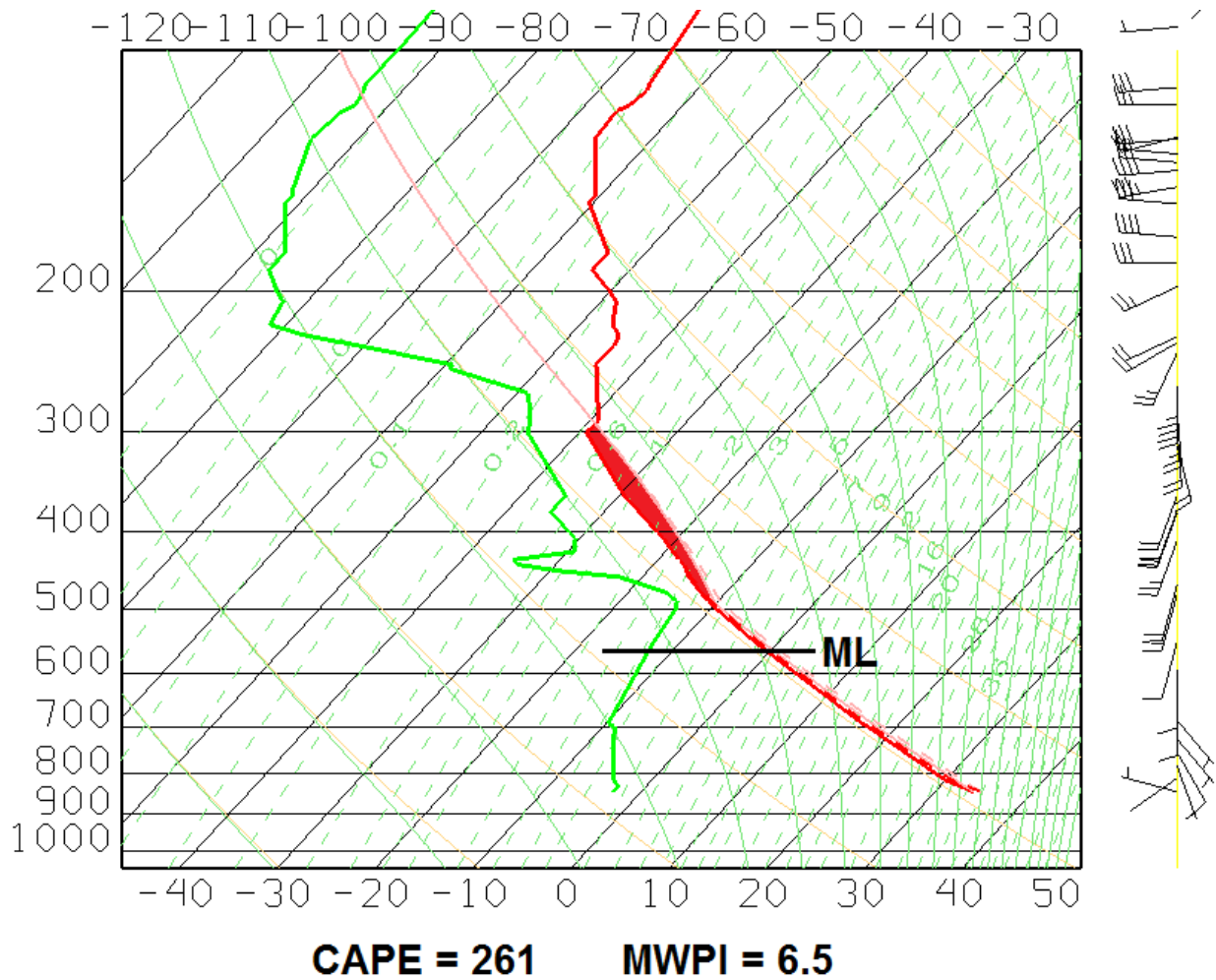

**CAPE = 261    MWPI = 6.5**

Figure 3. RAOB sounding thermodynamic profile over Reno, Nevada at 0000 UTC 2 July 2014. The dashed curve in thermodynamic profile represents virtual temperature. "ML" marks the height of the melting level (0°C isotherm). Mean layer CAPE, represented by the shaded area, is in units of J kg$^{-1}$. Vertical axis is labeled in units of pressure (millibars) and horizontal axis is labeled in units of temperature (°C).

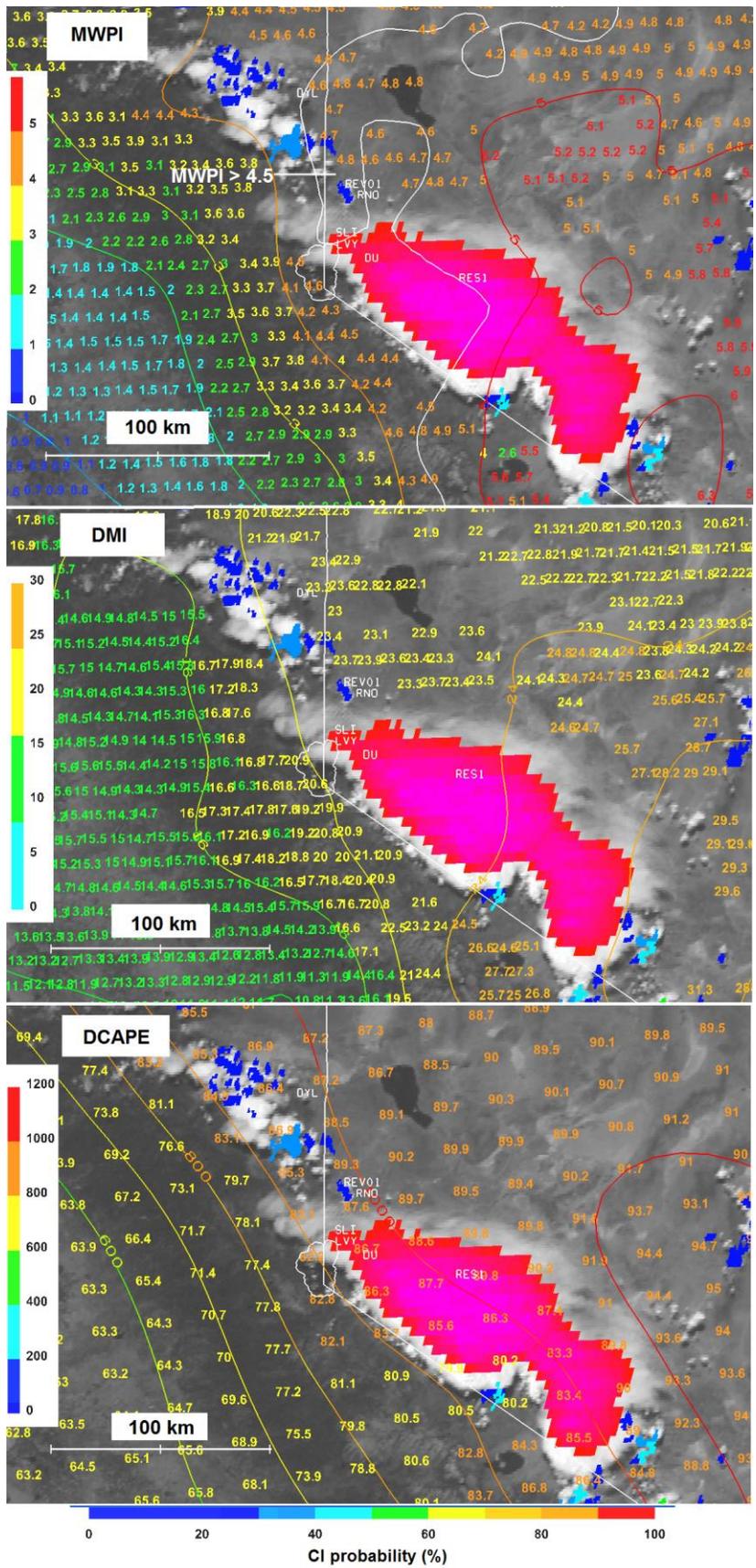

Figure 4. Comparison of GOES MWPI product (top), DMI product (middle), and RAP model DCAPE product (bottom) at 2200 UTC 1 July 2014. Contours of grid-interpolated MWPI and DMI values are overlying index values plotted at sounding retrieval locations and GOES-15 visible imagery (top, middle). Contours of grid-interpolated DCAPE values are overlying $W_{max}$ values (kt) plotted at RAP model grid points and GOES-15 visible imagery (bottom). The DMI contour interval is four (4) and the DCAPE contour interval is 100. Overlying BTD > −10°C is represented by the red to magenta color shading. Location identifiers of downburst occurrence are listed in Table 2.

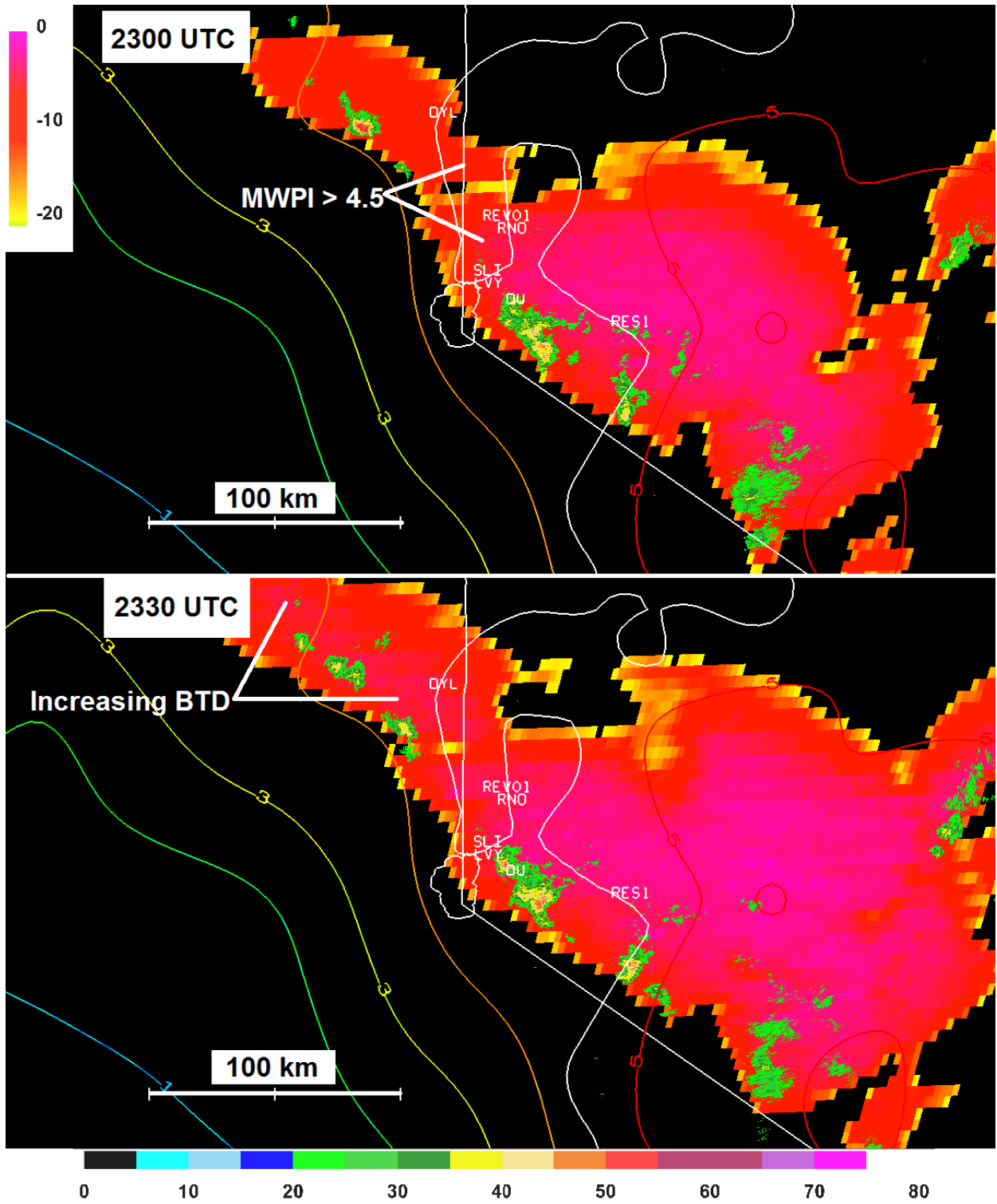

Figure 5. Blended GOES-15 imager WV-IR BTD and MWPI (color contours) product, with overlying 0.5° elevation radar reflectivity factor (dBZ) from Reno, Nevada NEXRAD at at 2300 UTC (top) and 2330 UTC (bottom) 1 July 2014. Color bars for BTD are located in the upper-right corner of the top product image; radar reflectivity factor color bar is located at the bottom of the figure.

Figure 6. Reno, Nevada 0.5° elevation NEXRAD reflectivity factor (dBZ) at 2330 UTC (top) and 2334 UTC (bottom) 1 July 2014 with 2200 UTC GOES-15 MWPI values plotted over the images. Location identifiers of downburst occurrence are listed in Table 2. "DU" marks the location of the report of a dust storm near Carson City at 2335 UTC. MWPI color bar is shown in Figure 4.

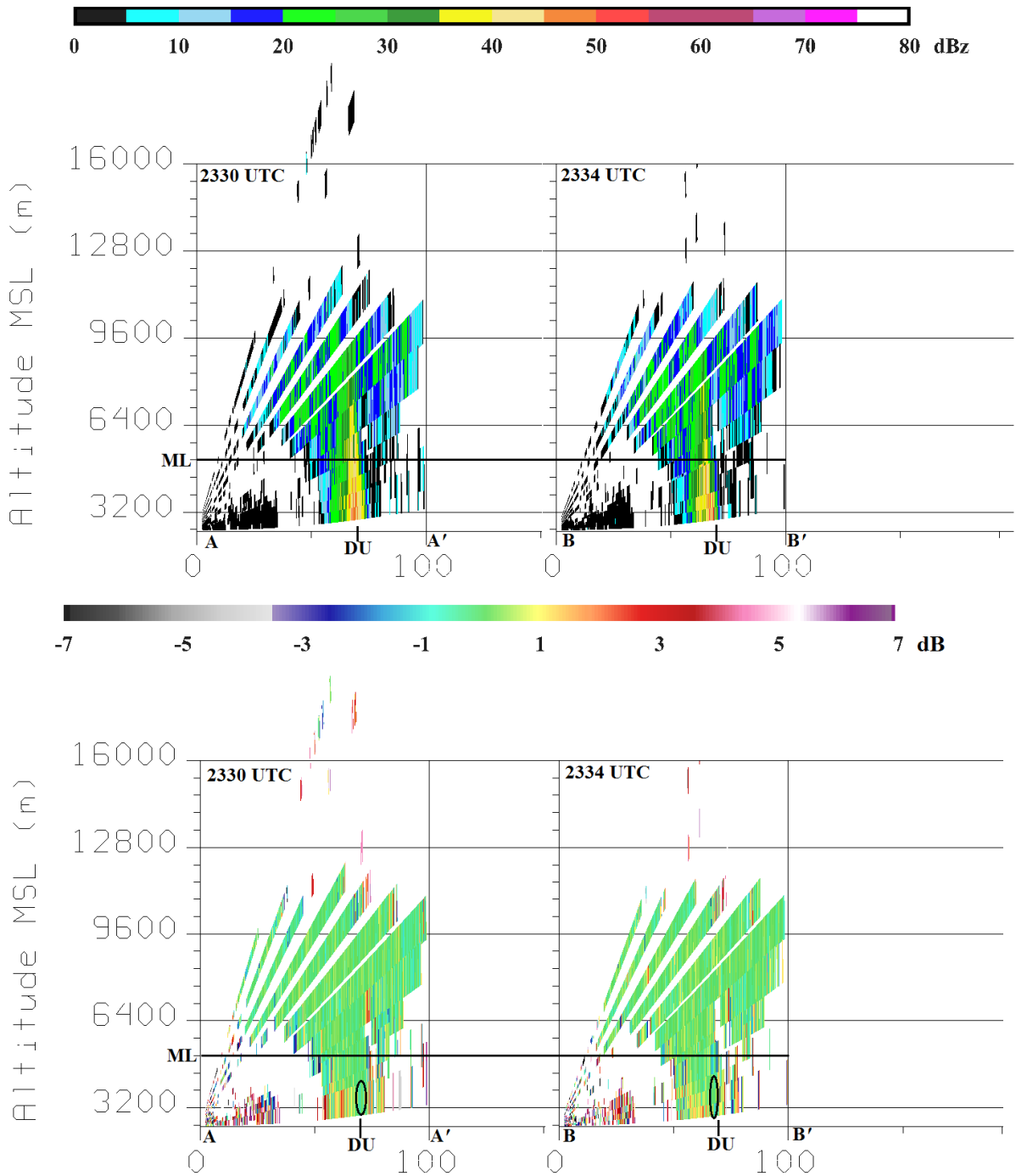

Figure 7. Reno, Nevada NEXRAD base reflectivity factor vertical transect images at 2330 UTC (top left) and 2334 UTC (top right) 1 July 2014; and differential reflectivity factor vertical transect images at 2330 UTC (bottom left) and 2334 UTC (bottom right) 1 July 2014. "DU" marks the location of the report of a dust storm near Carson City at 2335 UTC. "ML" represents

the height of the melting level indicated in the thermodynamic profile shown in Figure 3. Black-circled region indicates the presence of a hail core near the surface associated with a "$Z_{DR}$–hole" signature.

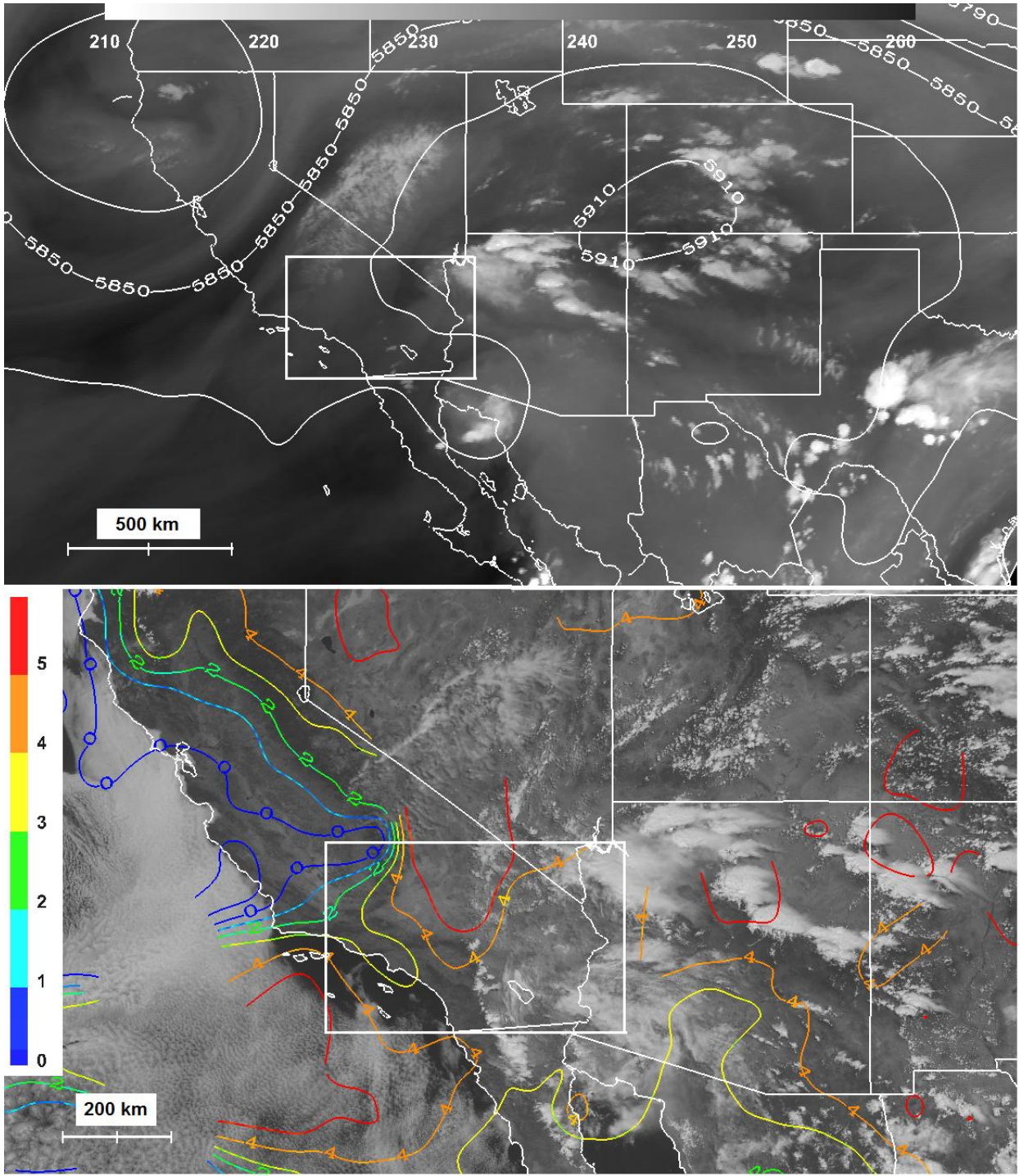

Figure 8. GOES-15 water vapor channel image with overlying 500 mb geopotential height contours (in meters, top) and GOES-15 MWPI product image (bottom) at 2000 UTC 18 August

2014. Contour interval is one (1) with contours for MWPI values of zero (0), two(2), and four (4) labeled. White rectangular area shows the domain of the images in Figure 10.

**CAPE = 180**

**CAPE = 2025**

Figure 9. RAP model-derived sounding thermodynamic profile over Daggett at 2000 UTC (top) and over Twenty-Nine Palms at 2100 UTC (bottom) 18 August 2014. The dashed curve in thermodynamic profile represents virtual temperature. "ML" marks the height of the melting level (0°C isotherm). Mean layer CAPE, represented by the shaded area, is in units of J kg$^{-1}$. Vertical axis is labeled in units of pressure (millibars) and horizontal axis is labeled in units of temperature (°C).

Figure 10. Comparison of GOES MWPI product (top), DMI product (middle), and RAP model DCAPE product (bottom) at 2000 UTC 18 August 2014. Contours of grid-interpolated MWPI and DMI values are overlying index values plotted at sounding retrieval locations and GOES-15 visible imagery (top, middle). Contours of grid-interpolated DCAPE values are overlying $W_{max}$ values (kt) plotted at RAP model grid points and GOES-15 visible imagery (bottom). The DMI contour interval is four (4) and the DCAPE contour interval is 100. "BYS", "DAG", "LAX, and "NXP" represent the locations of Fort Irwin, Daggett, Twenty-Nine Palms, and Los Angeles, California, respectively.

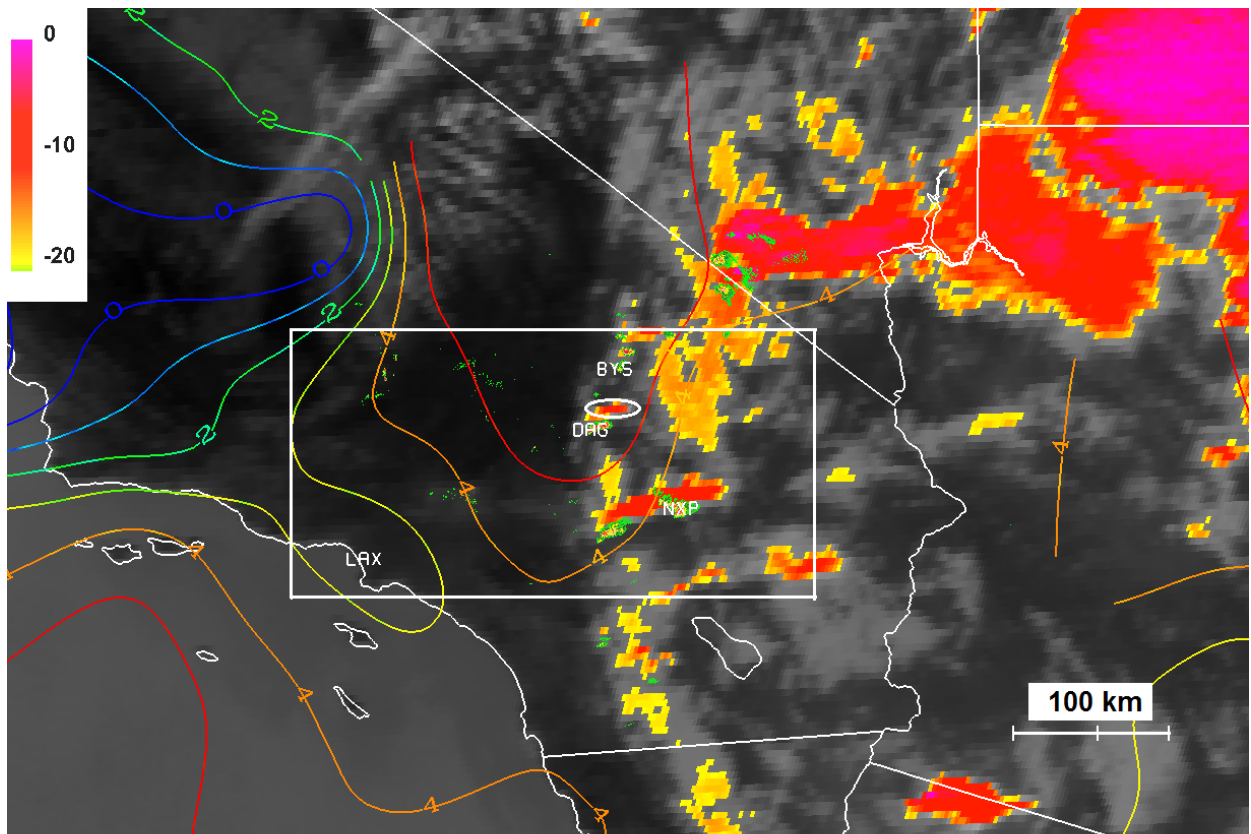

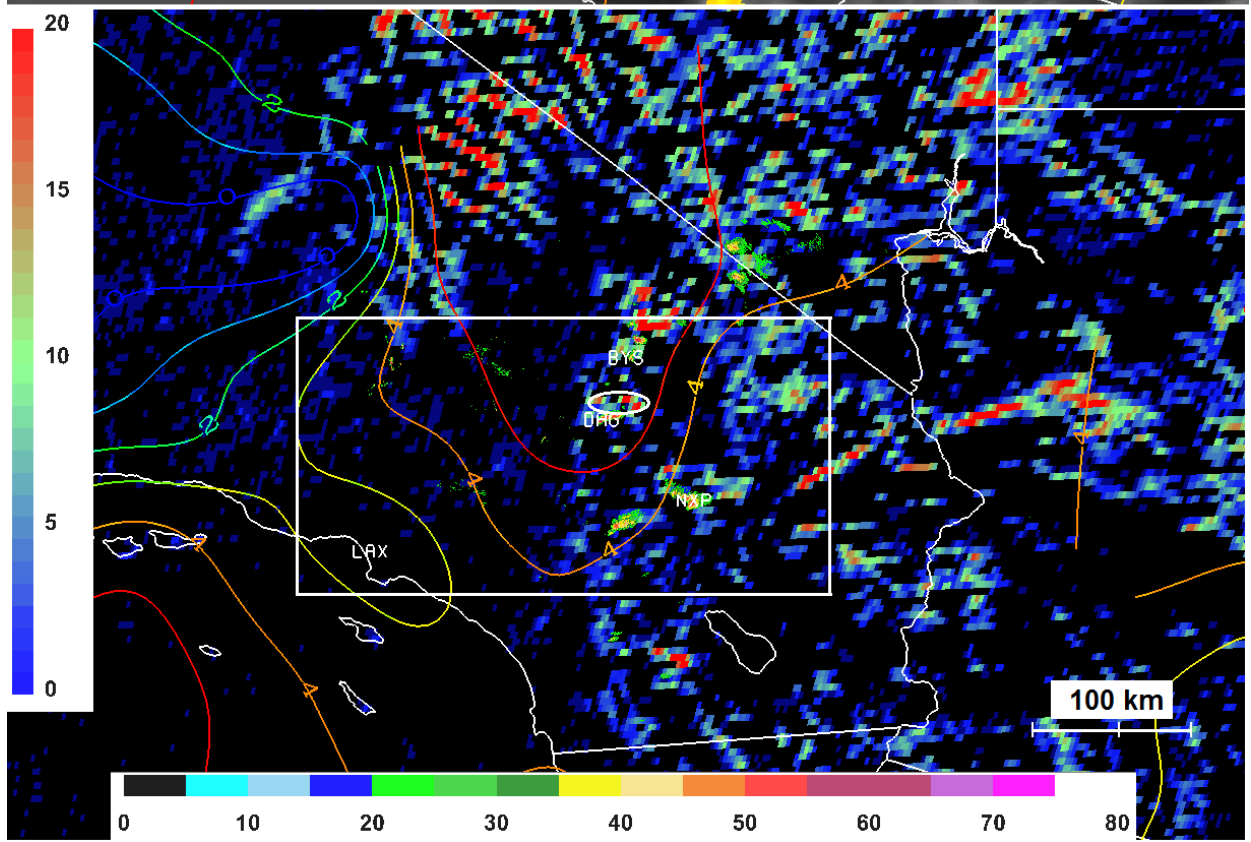

Figure 11. Blended GOES-15 imager WV-IR BTD and MWPI (color contours) product (top) and blended 11-minute ΔBTD and MWPI product (bottom) over southern California at 2211 UTC 18 August 2014, with overlying 0.5° elevation radar reflectivity factor (dBZ) from Edwards Air Force Base NEXRAD at 2212 UTC. Color bars for BTD and ΔBTD are located on the left hand side of the product image, radar reflectivity factor color bar is located at the bottom of the figure. "BYS", "DAG", "LAX, and "NXP" represent the locations of Fort Irwin, Daggett, Twenty-Nine Palms, and Los Angeles, California, respectively.

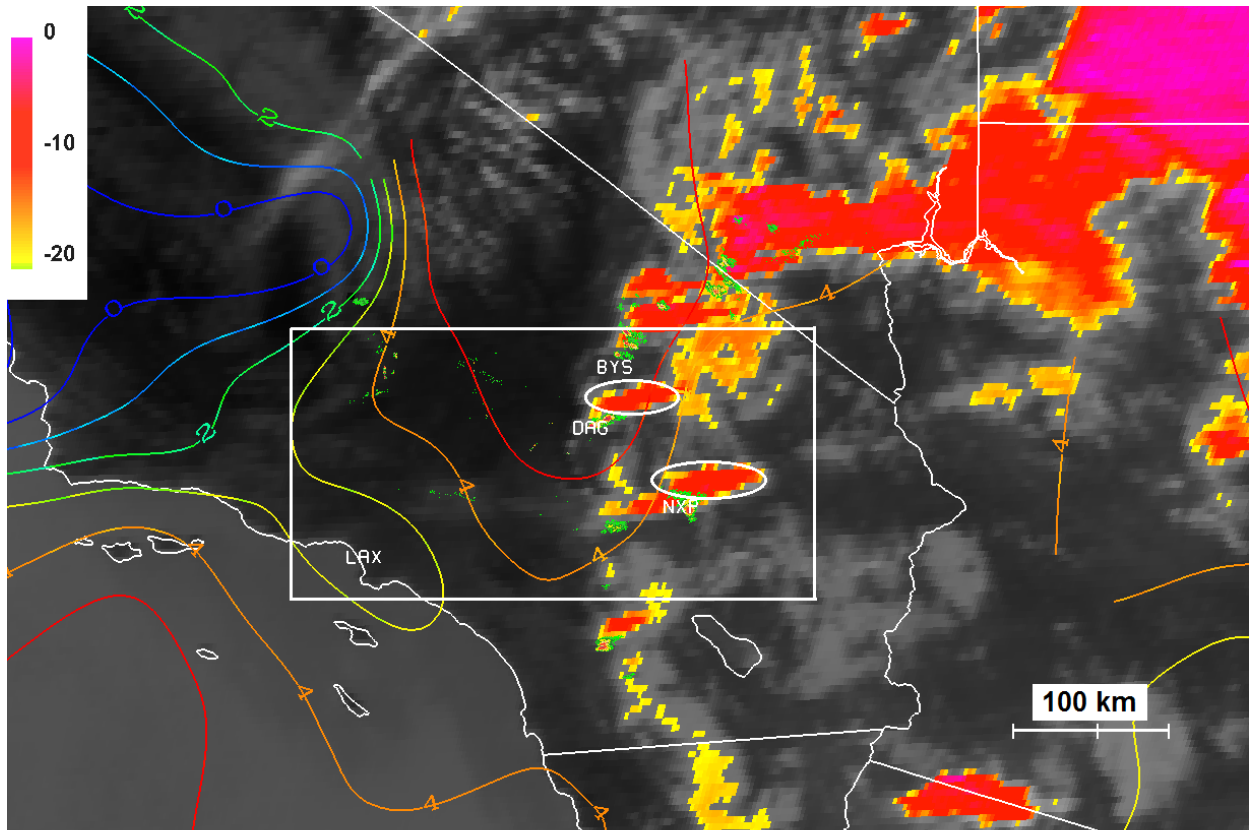

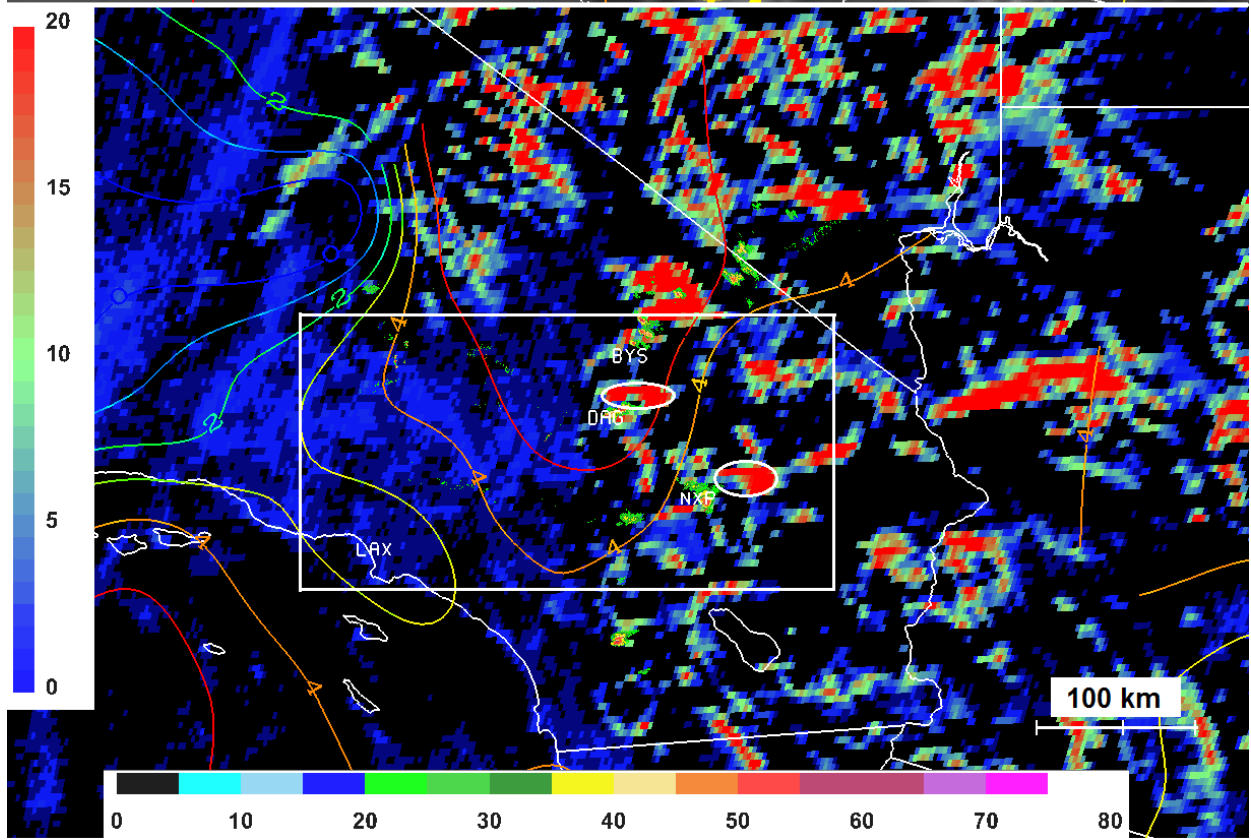

Figure 12. Blended GOES-15 imager WV-IR BTD and MWPI (color contours) product at 2241 UTC (top) and blended 19-minute ΔBTD and MWPI product (bottom) over southern California at 2230 UTC 18 August 2014, with overlying 0.5° elevation radar reflectivity factor (dBZ) from Edwards Air Force Base NEXRAD at 2236 UTC. Color bars for BTD and ΔBTD are located on the left hand side of the product image, radar reflectivity factor color bar is located at the bottom of the figure. "BYS", "DAG", "LAX, and "NXP" represent the locations of Fort Irwin, Daggett, Twenty-Nine Palms, and Los Angeles, California, respectively.

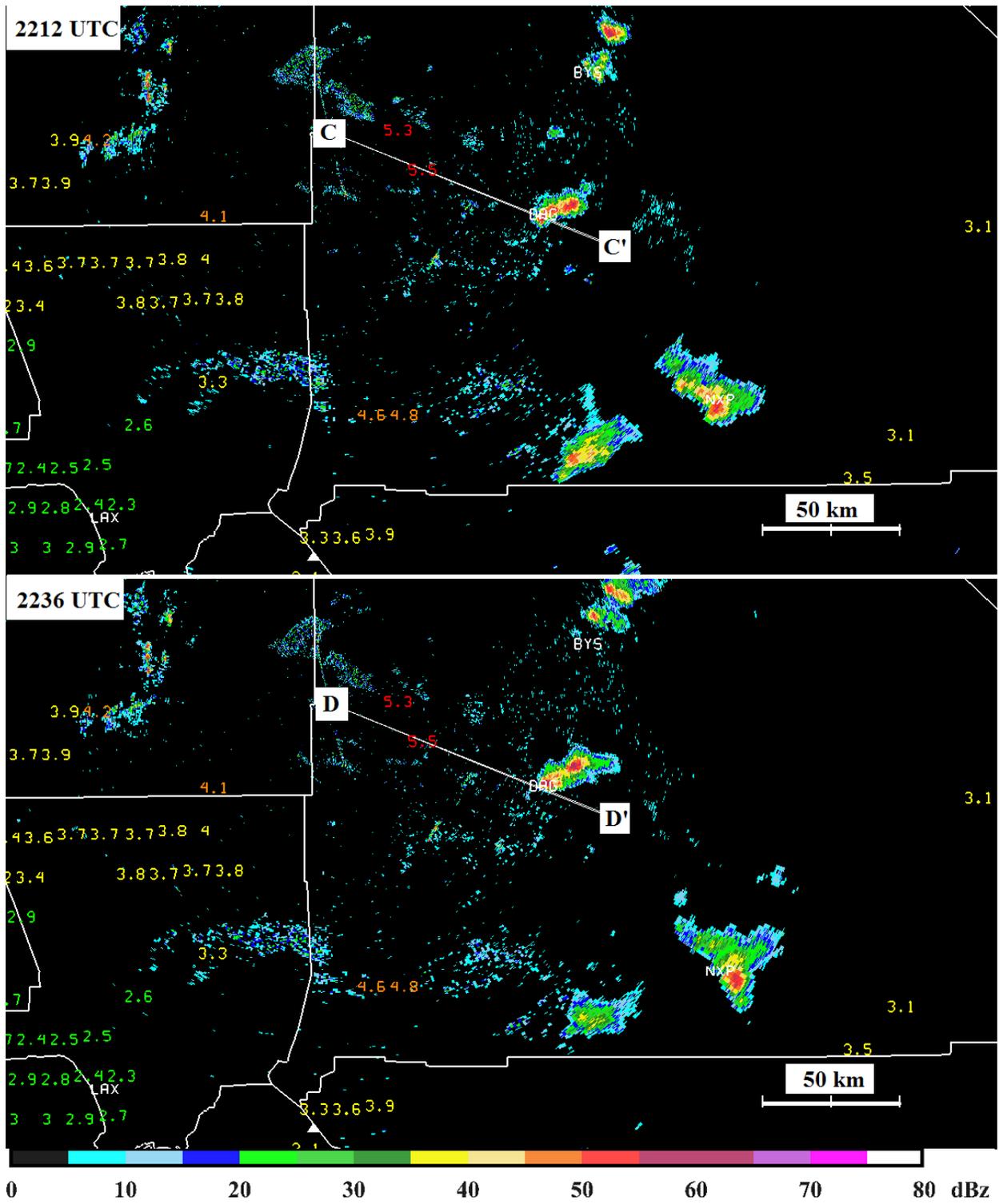

Figure 13. Edwards AFB, California 0.5° elevation NEXRAD reflectivity factor (dBZ) at 2212 UTC (top) and 2236 UTC (bottom) 18 August 2014 with 2000 UTC GOES-15 MWPI values plotted over the images. MWPI color scale is shown in Figure 10.  "BYS", "DAG", "LAX, and "NXP" represent the locations of Fort Irwin, Daggett, Twenty-Nine Palms, and Los Angeles, California, respectively.

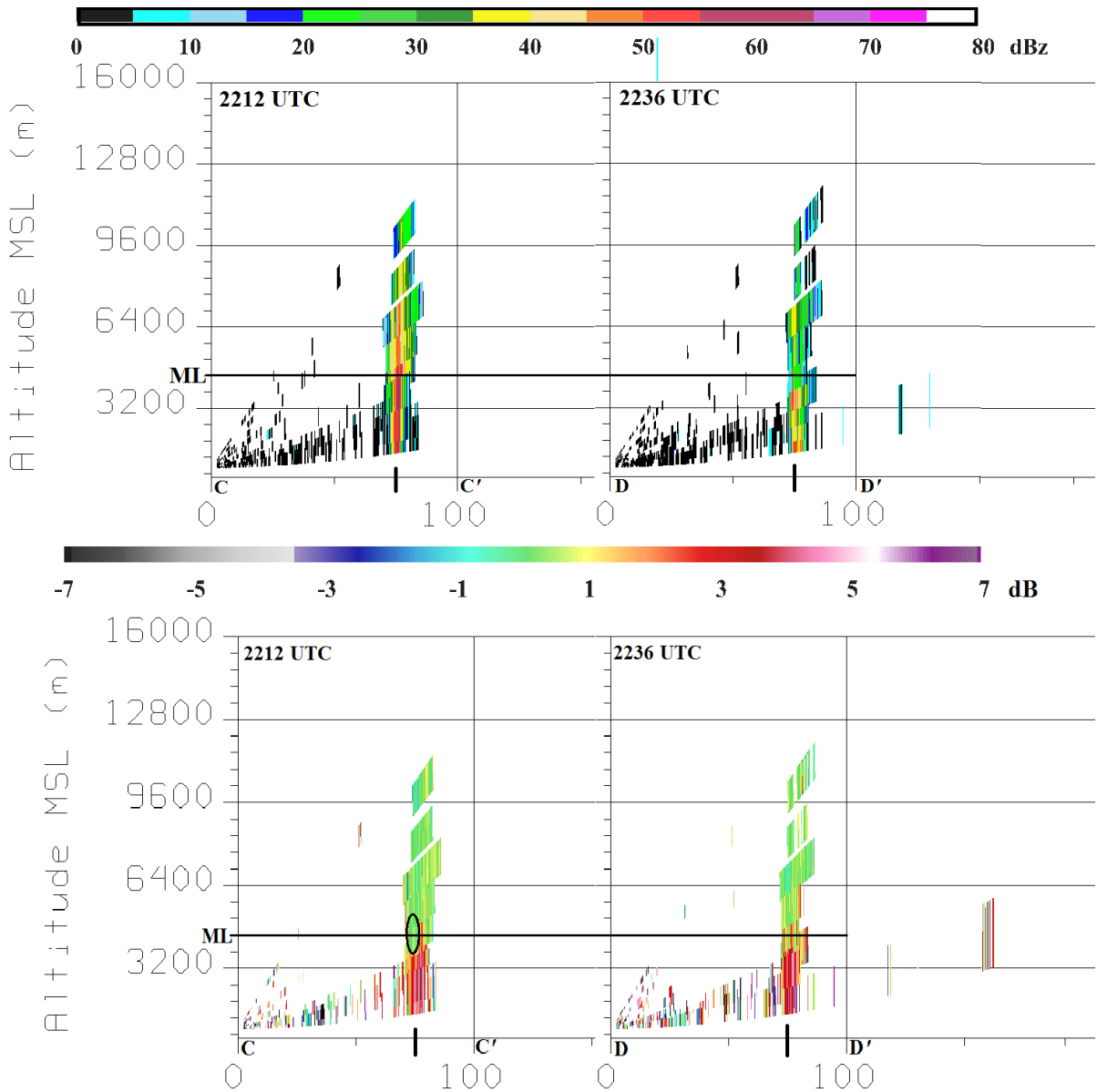

Figure 14. Edwards AFB, California NEXRAD base reflectivity factor vertical transect images at 2212 UTC (top left) and 2236 UTC (top right) 18 August 2014; and NEXRAD differential reflectivity factor vertical transect images at 2212 UTC (bottom left) and 2236 UTC (bottom right) 18 August 2014. Black tick marks the location of Barstow-Daggett Airport. "ML" represents the height of the melting level indicated in the thermodynamic profile shown in Figure 9. Black-circled region indicates the presence of a hail core near the melting level.